\documentclass[11pt,a4paper]{article}
\pdfoutput=1
\usepackage{jheppub}

\usepackage{color}
\usepackage{amsmath}
\usepackage{pifont}   
\usepackage{verbatim}       
\usepackage{acronym} 
    
\usepackage{amsfonts}
\usepackage{amssymb}
\usepackage{mathrsfs}
\usepackage{mathtools}
\usepackage{graphicx}
\usepackage{multirow} 
\usepackage{slashed} 
\usepackage{array}

\usepackage{graphicx}
\usepackage{epsfig}
\usepackage{lscape}
\usepackage{rotating}
\usepackage{epsf}
\usepackage{bm}
\usepackage{booktabs}
\usepackage{array}
\usepackage{caption}
\usepackage{subcaption}
\usepackage[utf8]{inputenc}
\usepackage{float}

\usepackage{multirow}
\usepackage{array,booktabs,colortbl,multirow}
\usepackage{colortbl,xcolor,colordvi,color}
\usepackage{rotating}
\allowdisplaybreaks

\usepackage{acronym}

\newcommand{\doublehat}[1]{\widehat{\widehat{#1}}}

\usepackage{multicol}

\title{Indirect detection constraints on semi-annihilation of inert scalar multiplets}

\author[a]{Hugues Beauchesne}
\author[b,a]{and Cheng-Wei Chiang}

\affiliation[a]{Physics Division, National Center for Theoretical Sciences, National Taiwan University,\\ Taipei 106319, Taiwan}
\affiliation[b]{Department of Physics and Center for Theoretical Physics, National Taiwan University, \\ Taipei 10617, Taiwan}

\emailAdd{beauchesneh@phys.ncts.ntu.edu.tw, chengwei@phys.ntu.edu.tw}

\abstract{Certain models of inert multiplets allow for semi-annihilation processes, in which two dark matter candidates annihilate to a dark matter particle and a non-dark matter particle. The existence of these processes can alleviate certain constraints and substantially modify the indirect detection signal. In this paper, we study current indirect detection constraints on the semi-annihilation of inert scalar multiplets. We show that there exist gauge numbers for which dark matter can be thermally produced and be compatible with indirect detection constraints even for very cuspy galactic dark matter density profiles.}

\begin{document}

\maketitle

%%%%%%%%%%%%%%%%%%%%%%%%%%%%%%%%%%%%%%%%%%%%%%%%%%
\section{Introduction}\label{Sec:Intro}
%%%%%%%%%%%%%%%%%%%%%%%%%%%%%%%%%%%%%%%%%%%%%%%%%%
Inert scalar multiplets are amongst the most studied dark matter candidates. These consist of scalar fields that transform under some representation of the electroweak gauge groups and do not acquire any expectation value nor mix with the Higgs. Previous works on inert multiplets include Refs.~\cite{Cirelli:2005uq, Cirelli:2007xd, Gustafsson:2007pc, Hambye:2009pw, Araki:2011hm, Fischer:2011zz, Josse-Michaux:2012oqz, Cai:2012kt, AbdusSalam:2013eya, Cheung:2013dua, Earl:2013jsa, Earl:2013fpa, Fischer:2013hwa, YaserAyazi:2014jby, Cai:2015kpa, Garcia-Cely:2015dda, Garcia-Cely:2015khw, Chowdhury:2016mtl, Kakizaki:2016dza, Khan:2016sxm, Lu:2016dbc, Logan:2016ivc, Cai:2017fmr, Cai:2017wdu, Liu:2017gfg, Cai:2018nob, Chao:2018xwz, Filimonova:2018qdc, Kadota:2018lrt, Zeng:2019tlw, Liu:2020dok, Jangid:2020qgo, Jueid:2020rek, Bottaro:2021snn, Bottaro:2022one}.

Even though they can naturally explain the observed dark matter abundance, inert multiplets are becoming increasingly constrained, namely by direct and indirect detection experiments. One way to suppress these constraints is via the inclusion of semi-annihilation processes~\cite{DEramo:2010keq}, which are possible when the model respects certain symmetries beyond the common $\mathbb{Z}_2$. A semi-annihilation process is a 2-to-2 collision in which two dark matter candidates annihilate to a dark matter candidate and another particle that is not part of the dark matter content. Such processes do not contribute to the direct detection signal at tree level and can have large cross sections. This allows for more efficient dark matter depletion, thus heavier dark matter and therefore a suppression of many experimental constraints. Previous works on semi-annihilation include Refs.~\cite{Hambye:2008bq, Batell:2010bp, Belanger:2012vp, Ivanov:2012hc, Belanger:2014bga, Cai:2016hne, Arcadi:2017vis, Kamada:2017gfc, Queiroz:2019acr, Yaguna:2019cvp, Ghosh:2020lma, Belanger:2020hyh, Belanger:2021lwd, Yaguna:2021rds, DiazSaez:2022nhp, Belanger:2022qxt, Bandyopadhyay:2022tsf, Cheng:2022hcm, BasiBeneito:2022qxd, Guo:2023kqt, Benincasa:2023vyp, Guo:2024jdj, Dominguez:2024gxh}.

In Ref.~\cite{Beauchesne:2024vbo}, we performed a detailed study of the possible models of inert multiplets semi-annihilation containing at most two multiplets. For models containing only one multiplet, it was shown that there does not exist any model in which semi-annihilation can have a significant impact on the dark matter abundance. For models containing two multiplets, several models can lead to efficient semi-annihilation, but there only exists one for which all dark matter subcomponents can give a suppressed direct detection signal and can avoid the existence of stable charged particles at tree level.

In the present paper, we continue the work of Ref.~\cite{Beauchesne:2024vbo} by studying current indirect detection constraints on inert multiplet semi-annihilation. Previous works on the indirect detection constraints on semi-annihilation include Refs.~\cite{Cai:2016hne, Queiroz:2019acr, Guo:2024jdj}.

We find the following results. Semi-annihilation can explain the observed dark matter abundance for all combinations of gauge numbers considered, as long as the galactic dark matter density profile has a sufficiently large core. The case of a triplet and a quadruplet is special, as semi-annihilation processes are Sommerfeld-suppressed. In this case, the correct dark matter abundance can be obtained even in the case of a very cuspy density profile.

The paper is organized as follows. The model and its attractive features are summarized in Sec.~\ref{Sec:Model}. Our implementation of the Sommerfeld enhancement is presented in Sec.~\ref{Sec:SE}. The particle spectra resulting from dark matter collisions are discussed in Sec.~\ref{Sec:Spectra}. Indirect detection constraints are described in Sec.~\ref{Sec:EC}. Sec.~\ref{Sec:Results} presents the results. Concluding remarks are given in Sec.~\ref{Sec:Conclusion}. Appendix~\ref{Sec:SU2tensors} contains the technical details of the $SU(2)$ tensors. Direct detection constraints are discussed in Appendix~\ref{Sec:DD}.

%%%%%%%%%%%%%%%%%%%%%%%%%%%%%%%%%%%%%%%%%%%%%%%%%%
\section{Model}\label{Sec:Model}
%%%%%%%%%%%%%%%%%%%%%%%%%%%%%%%%%%%%%%%%%%%%%%%%%%
We begin by presenting the model on which the analysis will focus. We follow closely Ref.~\cite{Beauchesne:2024vbo} and also include a few useful comments.

Introduce two complex scalar multiplets $\phi_1$ and $\phi_2$ of dimensions $n_1$ and $n_2$, respectively. We will always take $n_1$ odd and $n_2$ even. Unless stated otherwise, we will assume that $|n_1 - n_2| = 1$. The weak hypercharge of $\phi_1$ and $\phi_2$ are $Y_1 = 0$ and $Y_2 = -1/2$, respectively. The potential is given by
\begin{equation}\label{eq:Potential1A}
    V = V_0 + V_A + V_B + V_C,
\end{equation}
where
\begin{equation}\label{eq:Potential1B}
  \begin{aligned}
    V_0 =& -\mu^2 |H|^2 + \lambda_0 |H|^4 + m_1^2 |\phi_1|^2 + m_2^2 |\phi_2|^2,\\
    V_A =& \lambda_1 A_{abcd} \phi_1^a \phi_1^b \phi_2^c H^d + \lambda_2 B_{abc}\phi_1^a \phi_1^b \phi_1^c + \text{h.c.},\\
    V_B =& \sum_{r=1}^\alpha \lambda_3^r C^r_{abcd} (H^a)^\dagger H^b (\phi_1^c)^\dagger \phi_1^d + \sum_{r=1}^\beta \lambda_4^r D^r_{abcd} (H^a)^\dagger H^b (\phi_2^c)^\dagger \phi_2^d\\
         & + [\lambda_5 E_{abc} \phi_1^a (\phi_2^b)^\dagger (H^c)^\dagger + \text{h.c.}],\\
    V_C =& \sum_{r=1}^\gamma \lambda_6^r F^r_{abcd} (\phi_1^a)^\dagger (\phi_1^b)^\dagger \phi_1^c \phi_1^d + \sum_{r=1}^\delta \lambda_7^r G^r_{abcd} (\phi_2^a)^\dagger (\phi_2^b)^\dagger \phi_2^c \phi_2^d\\
         & + \sum_{r=1}^\epsilon \lambda_8^r H^r_{abcd} (\phi_1^a)^\dagger (\phi_2^b)^\dagger \phi_1^c \phi_2^d,
  \end{aligned}
\end{equation}
with $\lambda_{2,5}$ having dimensions of energy and all the other $\lambda$'s being dimensionless. Lowercase letters correspond to $SU(2)$ indices and are summed over. Uppercase letters correspond to $SU(2)$ tensors and their exact expressions are presented in Appendix~\ref{Sec:SU2tensors}. The following contractions of $SU(2)$ indices are non-vanishing when:
\begin{equation}\label{eq:TensorNonZero4}
  \begin{aligned}
    A: \quad & n_2 \in \{2n_1,\; 2n_1 - 2\;, 2n_1 - 4,\; ...\},\\
    B: \quad & n_1 \in \{1, 5, 9,\; ...\},\\
    E: \quad & n_1 - n_2 \in \{-1, 1\}.\\
  \end{aligned}
\end{equation}
All other contractions are always non-vanishing. The upper limits that appear in the sums of Eq.~\eqref{eq:Potential1B} are given by
\begin{equation}\label{eq:UpperLimits4}
  \begin{aligned}
    \alpha &= \text{min}\left(2, n_1\right), & \beta &= \text{min}\left(2, n_2\right), & \gamma &= \left\lfloor\frac{n_1 + 1}{2}\right\rfloor, &
    \delta &= \left\lfloor\frac{n_2 + 1}{2}\right\rfloor, & \epsilon &= \text{min}(n_1, n_2),
  \end{aligned}
\end{equation}
where $\lfloor x \rfloor$ means $x$ rounded down to the closest integer. The potential respects a $\mathbb{Z}_3$ symmetry under which
\begin{equation}\label{eq:SymmetryV4}
  \phi_1 \to e^{2\pi i/3}\phi_1, \quad \phi_2 \to e^{2\pi i/3}\phi_2.
\end{equation}
Once the Higgs acquires an expectation value, different components of $\phi_1$ and $\phi_2$ will mix, resulting in mass eigenstates $\hat{\phi}_i$. Unless $\mathbb{Z}_3$ is spontaneously broken, the lightest $\hat{\phi}_i$ will be stable. We will refer to it as $\hat{\phi}_0$. The masses of the $\hat{\phi}_i$ are labelled as $\hat{m}_i$ and the lightest one is referred to as $\hat{m}_0$.

The model possesses several appealing features. First, semi-annihilation processes can have cross sections much larger than those of annihilation to gauge bosons. This is a non-trivial property and cannot be satisfied with a single multiplet (see Ref.~\cite{Beauchesne:2024vbo} for a proof of this statement). Efficient semi-annihilation is possible because of the $\lambda_1$ term. Second, as long as $m_1 < m_2$ and the different $\lambda$'s are not too large, $\hat{\phi}_0$ will have almost no tree-level interactions with the $Z$ boson and can pass direct detection constraints even at loop level \cite{Chen:2023bwg}. We will therefore assume that $m_1 < m_2$ is respected throughout this work. Third, it is possible to choose values of $m_1$, $m_2$, $\lambda_3^2$ and $\lambda_5$ such that $\hat{\phi}_0$ is neutral even at the tree level.

%%%%%%%%%%%%%%%%%%%%%%%%%%%%%%%%%%%%%%%%%%%%%%%%%%
\section{Sommerfeld enhancement}\label{Sec:SE}
%%%%%%%%%%%%%%%%%%%%%%%%%%%%%%%%%%%%%%%%%%%%%%%%%%
When dark matter annihilates at low velocities, such as in the galactic center or in dwarf galaxies, the exchange of gauge or Higgs bosons can drastically modify the dark matter annihilation rate via the Sommerfeld enhancement~\cite{Sommerfeld:1931qaf}. In this section, we explain our implementation of this phenomenon. We will take inspiration from Ref.~\cite{Cirelli:2015bda}, from which many useful results can be extracted. We also refer to Refs.~\cite{Hisano:2004ds, Cirelli:2007xd, Arkani-Hamed:2008hhe, Slatyer:2009vg, Iengo:2009ni, Cassel:2009wt} for proofs and more detailed explanations. For the sake of keeping the computations manageable, we will assume that the mass eigenstates corresponding mostly to $\phi_2$ can be integrated out. This should be an excellent approximation unless $m_2/m_1$ is very close to 1 and $\lambda_5$ large~\cite{Beneke:2014gja}.

\subsection{Computation of the Sommerfeld factors}\label{sSec:SEFactors}
We begin by explaining our procedure for solving the Schr\"odinger equation.

We will be interested in the gauge and Higgs interactions of the $n_1$ lightest mass eigenstates, which correspond mostly to the components of $\phi_1$. For convenience's sake, assume the light $\hat{\phi}_i$ are sorted from most positive to most negative electric charge. The gauge and Higgs interactions can be parametrized following the notation of Ref.~\cite{Beauchesne:2023iyn} as:
\begin{equation}\label{eq:LagrangianGaugeCS}
  \begin{aligned}
    \mathcal{L}^{\text{light}}= \;\;\;& -i \hat{A}_{ii} A_\mu \left({\hat{\phi}_i}^\dagger \partial^\mu \hat{\phi}_i - \partial^\mu \hat{\phi}_i^\dagger \hat{\phi}_i\right)
                                        -i \hat{C}_{ij} Z_\mu \left({\hat{\phi}_i}^\dagger \partial^\mu \hat{\phi}_j - \partial^\mu \hat{\phi}_j^\dagger \hat{\phi}_i\right)\\
                                      & -\left[i \hat{F}_{ij}W^+_\mu\left({\hat{\phi}_i}^\dagger \partial^\mu \hat{\phi}_j - \partial^\mu{\hat{\phi}_i}^\dagger \hat{\phi}_j\right) + \text{h.c.}\right]
                                        -\hat{\Omega}_{ij} h \hat{\phi}_i^\dagger \hat{\phi}_j.
  \end{aligned}
\end{equation}
In practice, $\hat{A}$ is simply $\hat{A} = e T_3$, with $T_3$ being the third $SU(2)$ generator of dimension $n_1$. The $\hat{C}$, $\hat{F}$ and $\hat{\Omega}$ matrices can, however, be slightly more complicated because of mixing.

When two dark matter particles approach each other, the two particles are at first neutral, but the exchange of $W$ bosons will turn the wavefunction into a linear combination of different pairs of particles of opposite charges. Consider the scattering $\hat{\phi}_i^\dagger \hat{\phi}_i \to \hat{\phi}_j^\dagger \hat{\phi}_j$. The corresponding $n_1 \times n_1$ potential is
\begin{equation}\label{eq:PotentialPA}
  V_{ij} = -\frac{\hat{A}_{ii}^2}{4\pi r}\delta_{ij} - \frac{\hat{C}_{ii}^2}{4\pi r}\delta_{ij}e^{-m_Z r} - \frac{\hat{F}_{ij}^2}{4\pi r}e^{-m_W r} - \frac{\hat{F}_{ji}^2}{4\pi r}e^{-m_W r} - \frac{\hat{\Omega}_{ii}^2}{16\pi \hat{m}_0^2r}\delta_{ij}e^{-m_h r} + \delta \hat{m}_{ij},
\end{equation}
where $\delta \hat{m}_{ij} = 2(\hat{m}_i - \hat{m}_0)\delta_{ij}$ and we have exploited the fact that $\hat{A}$, $\hat{C}$ and $\hat{\Omega}$ are diagonal when the mostly $\phi_2$ mass eigenstates are integrated out. The mass splitting includes effects from mixing and radiative corrections~\cite{Cirelli:2005uq}.

Similarly, consider the scattering $\hat{\phi}_i \hat{\phi}_{i'} \to \hat{\phi}_j \hat{\phi}_{j'}$, where $\hat{\phi}_{i'}$ is the mass eigenstate with opposite charge to $\hat{\phi}_i$ and $\hat{\phi}_{j'}$ is the mass eigenstate with opposite charge to $\hat{\phi}_j$. There are $n \equiv (n_1 + 1)/2$ possible particle pairs. The corresponding $n \times n$ potential is
\begin{equation}\label{eq:PotentialPP}
  \begin{aligned}
  \tilde{V}_{ij} &= \left[\frac{\hat{A}_{ii} \hat{A}_{i'i'}}{4\pi r}\delta_{ij} + \frac{\hat{C}_{ii} \hat{C}_{i'i'}}{4\pi r}\delta_{ij}e^{-m_Z r} + \frac{\hat{F}_{ij}\hat{F}_{j'i'}}{4\pi r}e^{-m_W r}\right.\\
                & \hspace{0.8cm}  + \frac{\hat{F}_{ji} \hat{F}_{i'j'}}{4\pi r}e^{-m_W r} - \left.\frac{\hat{\Omega}_{ii} \hat{\Omega}_{i'i'}}{16\pi \hat{m}_0^2 r}\delta_{ij}e^{-m_h r} + \delta \tilde{m}_{ij}\right] s_i s_j,
  \end{aligned}
\end{equation}
where $\delta\tilde{m}_{ij} = (\hat{m}_i + \hat{m}_{i'} - 2\hat{m}_0)\delta_{ij}$ and the factor $s_a = \sqrt{1 + \delta_{aa'}}$ accounts for the presence or absence of identical particles in the incoming or outgoing states.

To obtain the Sommerfeld factors, one must solve the following radial Schr\"odinger equations:
\begin{equation}\label{eq:radialSchrodinger}
  \begin{aligned}
    -\frac{1}{\hat{m}_0}\frac{d^2}{dr^2}         u_{ia}(r) +         V_{ij}         u_{ja}(r) &= E         u_{ia}(r),\\
    -\frac{1}{\hat{m}_0}\frac{d^2}{dr^2} \tilde{u}_{ia}(r) + \tilde{V}_{ij} \tilde{u}_{ja}(r) &= E \tilde{u}_{ia}(r).
  \end{aligned}
\end{equation}
The functions $u_a(r)$ represent the radial part of the wave function divided by the radius $r$. The subscript $a$ is the label of the solution and, since there are $n_1$ distinct physical solutions, it runs from 1 to $n_1$. Each element $u_{ia}(r)$ corresponds to the component $\hat{\phi}_i^\dagger \hat{\phi}_i$ of that state at that radius. The index $i$ runs over all combinations, i.e. from 1 to $n_1$. The functions $\tilde{u}_{ia}(r)$ are similar, but correspond to states $\hat{\phi}_i \hat{\phi}_{i'}$ (up to a $1/\sqrt{2}$ factor for identical particles) and $i$ and $a$ run from 1 to $n$. The energy is given by $E = \hat{m}_0 v^2/4$, where $v$ is the relative velocity of the two colliding particles. The boundary conditions are
\begin{equation}\label{eq:Boundaries}
  \begin{aligned}
    &         u_{ia}(0) = \delta_{ia}, \quad \frac{d}{dr}         u_{ia}(\infty) = i k_i                 u_{ia},\\
    & \tilde{u}_{ia}(0) = \delta_{ia}, \quad \frac{d}{dr} \tilde{u}_{ia}(\infty) = i \tilde{k}_i \tilde{u}_{ia},
  \end{aligned}
\end{equation}
where $k_i = \sqrt{\hat{m}_0(E - \delta\hat{m}_{ii})}$ and $\tilde{k}_i = \sqrt{\hat{m}_0(E - \delta\tilde{m}_{ii})}$. The matrices $R$ and $\tilde{R}$ are defined through the asymptotic behaviour of the wave function
\begin{equation}\label{eq:SEAsymptotic}
  u_{ia}(r \to \infty) = R_{ia} e^{ik_i r}, \quad \tilde{u}_{ia}(r \to \infty) = \tilde{R}_{ia} e^{i\tilde{k}_i r}.
\end{equation}
The matrices of Sommerfeld factors are then simply given by
\begin{equation}\label{eq:SEFactors}
  s = R^T, \quad \tilde{s} = \tilde{R}^T.
\end{equation}
In practice, the exchange of Higgs bosons is suppressed by factors of either $(v_H/m_1)^2$ or $|\lambda_5|^2/(m_2^2 - m_1^2)$, where $v_H$ is the Higgs vacuum expectation value. These factors must be small to reproduce the observed dark matter abundance and satisfy direct detection constraints. The Higgs exchange will therefore only have a small effect on the final results.

\subsection{Cross sections}\label{sSec:CrossSections}
Once the Sommerfeld factors have been computed, obtaining the cross sections is straightforward. During the annihilation process, one can neglect the effect of electroweak symmetry breaking on the $\hat{\phi}_i$. We will also take the limit of low velocity and heavy $\hat{\phi}_i$.

First, consider an annihilation process between a $\hat{\phi}_0^\dagger$ and a $\hat{\phi}_0$ to a final state $f$. The Sommerfeld-enhanced cross section is given by
\begin{equation}\label{eq:SEPACS}
  (\sigma v)_f = (s^\dagger \Gamma_f s)_{nn},
\end{equation}
where $\Gamma_f$ is the rate associated to the final state $f$. The rates for annihilation to gauge or Higgs bosons are given by
\begin{equation}\label{eq:GammaGaugeBosons}
  \begin{aligned}
     \Gamma_{Z Z}     &= \frac{g^4 c_W^4}{8 \pi m_1^2}\hat{T}_N \hat{T}_N^T + \frac{\hat{\lambda}_+ \hat{\lambda}^T_+}{64 \pi m_1^2}, \quad 
    &\Gamma_{Z A}     &= \frac{g^4 c_W^2 s_W^2}{4 \pi m_1^2}\hat{T}_N \hat{T}_N^T, \quad
    &\Gamma_{A A}     &= \frac{g^4 s_W^4}{8 \pi m_1^2}\hat{T}_N \hat{T}_N^T, \\
     \Gamma_{W^+ W^-} &= \frac{g^4}{64 \pi m_1^2}\hat{T}_C \hat{T}_C^T + \frac{\hat{\lambda}_- \hat{\lambda}_-^T}{32 \pi m_1^2}, \quad
    &\Gamma_{h h}     &= \frac{\hat{\lambda}_+ \hat{\lambda}_+^T}{64 \pi m_1^2},
  \end{aligned}
\end{equation}
where $g$ is the $SU(2)_L$ gauge coupling, $s_W$ ($c_W$) are the sine (cosine) of the weak angle $\theta_W$ and the column vectors $\hat{T}_N$ and $\hat{T}_C$ are defined as
\begin{equation}\label{eq:T3TC}
  (\hat{T}_N)_i = (T_3)_{ii}^2, \qquad (\hat{T}_C)_i = \left(\{T^+, T^-\} \right)_{ii},
\end{equation}
where $T^+_{ab} = T^-_{ba} = \sqrt{a(n_1 - a)}\delta_{a, b - 1}$. The column vector $\hat{\lambda}_\pm$ is defined as $\hat{\lambda}_\pm \equiv -\lambda_3^1 \pm \lambda_3^2 \hat{T}_3/2$, with $(\hat{T}_3)_i = (T_3)_{ii}$. The presence of the $\hat{\lambda}_\pm$ terms in the rates to gauge bosons is a consequence of the Goldstone boson equivalence theorem.

Second, consider an annihilation process of two $\hat{\phi}_0$ to a final state $f$. In practice, this is always a semi-annihilation process. The Sommerfeld-enhanced cross section is given by
\begin{equation}\label{eq:SEPPCS}
  (\sigma v)_f = 2 (\tilde{s}^\dagger \Gamma_f \tilde{s})_{nn},
\end{equation}
where $\Gamma_f$ is the rate associated to the final state $f$. The rate for semi-annihilation is
\begin{equation}\label{eq:GammaSA}
  \Gamma_{\tilde{\phi}^\dagger {W^-}} = 2\Gamma_{\tilde{\phi}^\dagger Z} = 2\Gamma_{\tilde{\phi}^\dagger h} = \frac{|\lambda_1|^2}{32\pi m_1^2}\left(1 - \frac{m_2^2}{4m_1^2}\right) \tilde{M}_{n_1 n_2},
\end{equation}
where $\tilde{\phi}$ corresponds to a $\hat{\phi}_i$ that is mostly $\phi_2$. For the cases we will be interested in, the $\tilde{M}_{n_1 n_2}$ matrices are
\begin{equation}\label{eq:Mtilden1n2}
  \begin{aligned}
    &\tilde{M}_{3 2} = \frac{1}{3}  \begin{pmatrix} 2 & -\sqrt{2} \\ -\sqrt{2} & 1 \end{pmatrix}, \;
     \tilde{M}_{3 4} = \frac{1}{15} \begin{pmatrix} 1 &  \sqrt{2} \\  \sqrt{2} & 2 \end{pmatrix},  \\
    &\tilde{M}_{5 4} = \tilde{M}_{5 6} = \frac{1}{35} \begin{pmatrix} 4 & 2 & -2\sqrt{2} \\ 2 & 1 & -\sqrt{2} \\ -2\sqrt{2} & -\sqrt{2} & 2 \end{pmatrix},\;
     \tilde{M}_{7 6} = \frac{1}{210} \begin{pmatrix} 25 & 0 & -15 & 10\sqrt{2} \\ 0 & 0 & 0 & 0 \\ -15 & 0 & 9 & -6\sqrt{2} \\ 10\sqrt{2} & 0 & -6\sqrt{2} & 8 \end{pmatrix}. 
  \end{aligned}
\end{equation}

\subsection{Illustration of the Sommerfeld enhancement}\label{sSec:SEIllustration}
\begin{figure}[t!]
    \centering
    \begin{subfigure}{0.49\textwidth}
        \centering
        \caption{}
        \includegraphics[width=\textwidth]{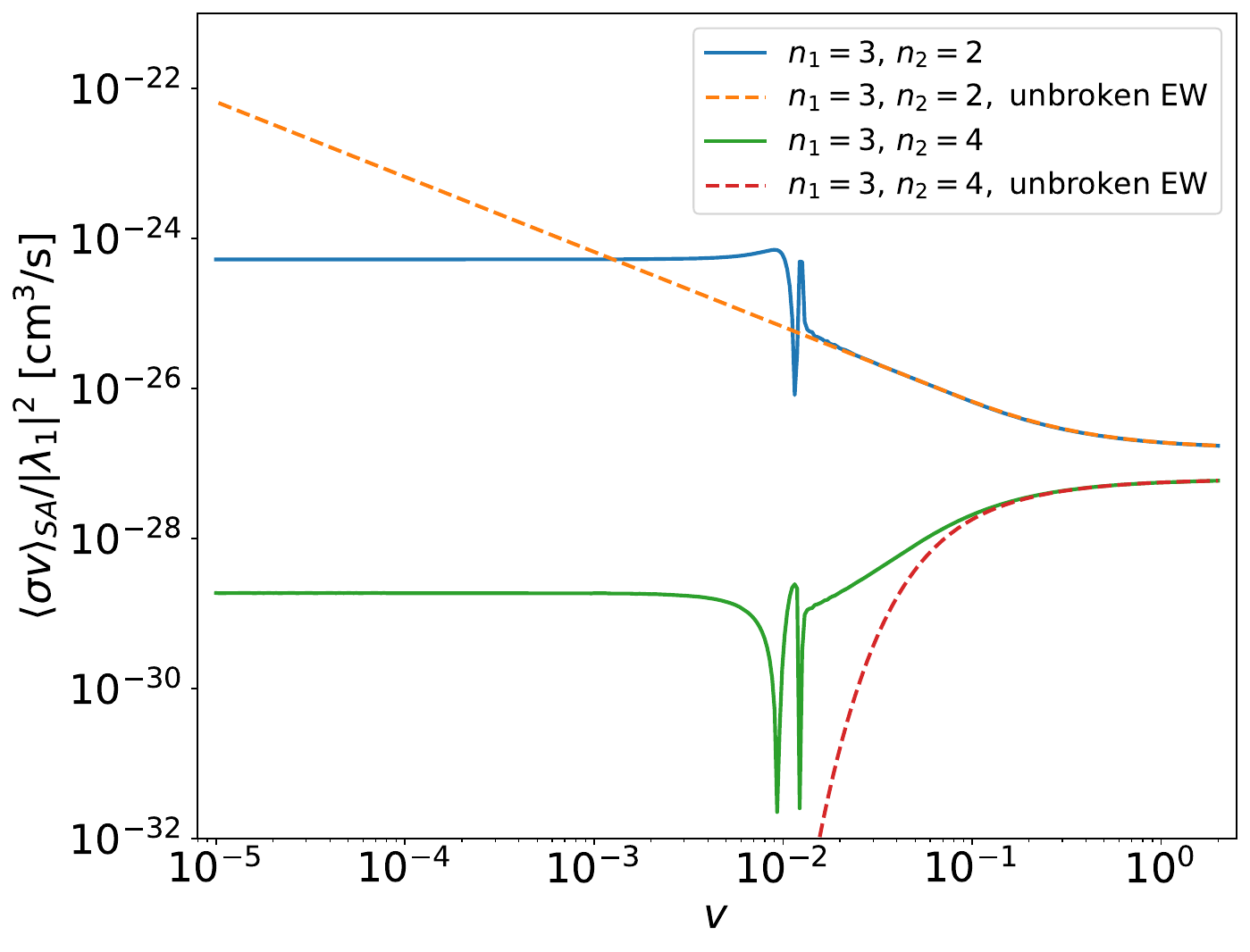}
        \label{fig:vDependence}
    \end{subfigure}    
    \begin{subfigure}{0.49\textwidth}
        \centering
        \caption{}
        \includegraphics[width=\textwidth]{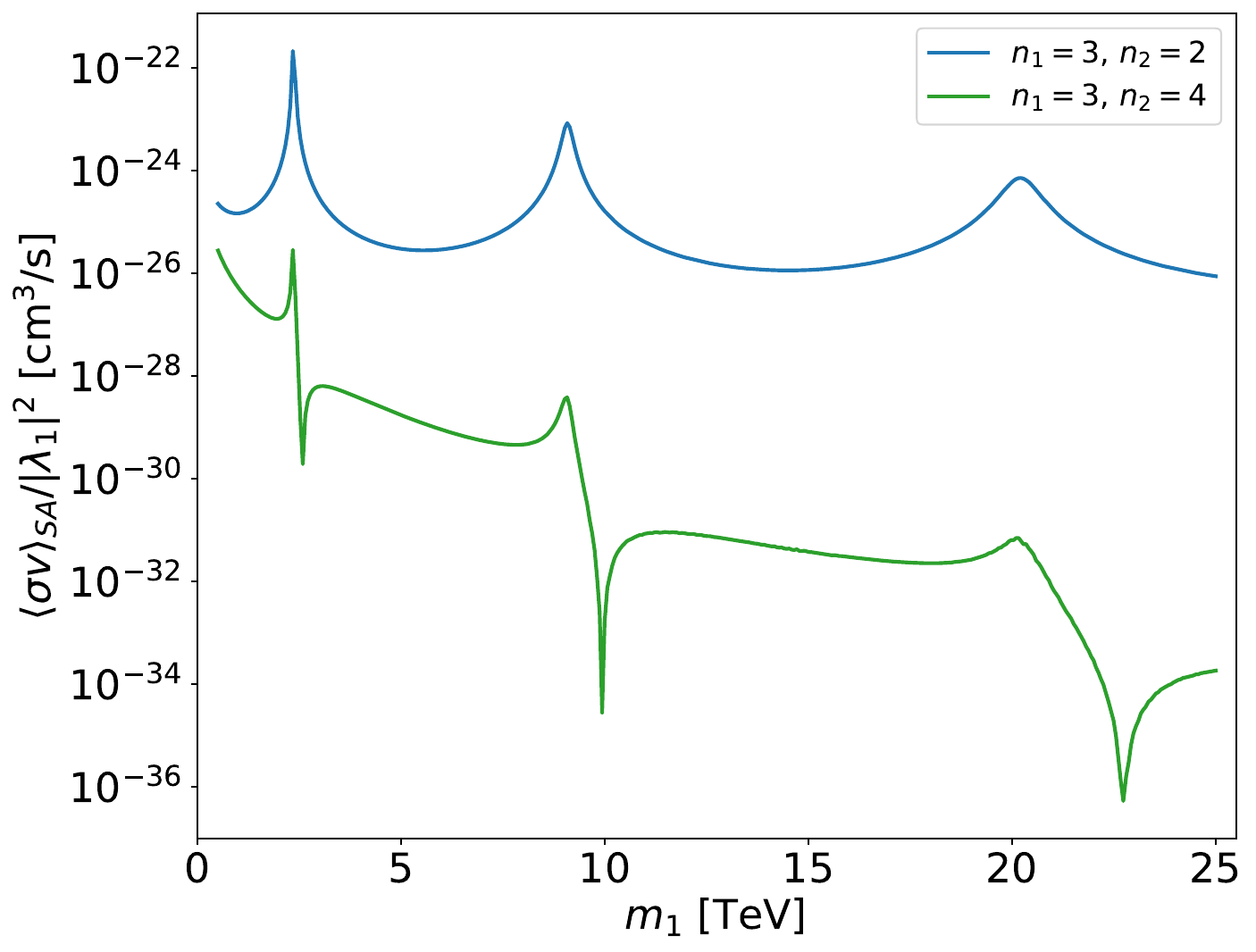}
        \label{fig:mDependence}
    \end{subfigure}    
    \caption{(a) Total semi-annihilation cross section as a function of the relative velocity and normalized by $|\lambda_1|^2$. The mass $m_1$ is set to 8~TeV. The analytical results of Eq.~\eqref{eq:SE} are also included. (b) Total semi-annihilation cross section as a function of the mass $m_1$ and normalized by $|\lambda_1|^2$. The relative velocity is set to $10^{-3}$. For both (a) and (b), $m_2/m_1 = 1.2$, all other parameters are set to 0 and the results are independent of $\lambda_1$.}
    \label{fig:SE}
\end{figure}

We show in Fig.~\ref{fig:SE} the effect of the Sommerfeld enhancement on the total semi-annihilation cross section for different choices of gauge numbers. As can be seen, the cross section is enhanced for $n_1 = 3$ and $n_2 = 2$, but suppressed for $n_1 = 3$ and $n_2 = 4$. All other combinations of $n_1$ and $n_2$ that satisfy $|n_1 - n_2| = 1$ are enhanced.

To understand this behaviour, one can look at the limit in which electroweak symmetries are restored, in which case an analytical solution is known. Consider two colliding particles $P_A$ and $P_B$ of weak isospins $J_A$ and $J_B$ and weak hypercharge $Y_A$ and $Y_B$, respectively. Assume they form a state of weak isospin $J$. Define~\cite{Strumia:2008cf}
\begin{equation}\label{eq:alpha}
  \alpha_{\text{SE}} = \frac{g^2}{8\pi}\left[J(J + 1) - J_A(J_A + 1) - J_B(J_B + 1) \right] + \frac{(g')^2}{4\pi}Y_A Y_B \eta,
\end{equation}
where $\eta = -1$ for the collision of a particle and an antiparticle and $\eta = 1$ otherwise. The Sommerfeld enhancement is then \cite{Harris:1957zza}
\begin{equation}\label{eq:SE}
  S_{\text{SE}} = -\frac{2\pi \alpha_{\text{SE}}}{v}\frac{1}{1 - e^{\frac{2\pi \alpha_{\text{SE}}}{v}}}.
\end{equation}
For semi-annihilation of two $\phi_1$ and $|n_1 - n_2| = 1$, $J$ is given by $J = 2\lfloor n_2/4 \rfloor$. It is a simple task to verify that $\alpha_{\text{SE}}$ is negative for every choice of $n_1$ and $n_2$ that satisfies $|n_1 - n_2| = 1$ except $n_1 = 3$ and $n_2 = 4$. This is why $n_1 = 3$ and $n_2 = 4$ is suppressed by the Sommerfeld effect. This suppression will result in the indirect detection signal of semi-annihilation being negligible for this choice of gauge numbers, which will help alleviate constraints.

Fig.~\ref{fig:SE} also contains the analytical solutions of Eq.~\eqref{eq:SE}. As can be seen, they match to excellent precision at high velocity. As the velocity decreases, the masses of the gauge bosons cannot be neglected anymore and the approximation breaks down as expected. The Sommerfeld enhancement then reaches an asymptotic behaviour. When discussing indirect detection constraints, we will, in practice, take a relative velocity of $v = 10^{-3}$ for the galactic center and $v = 3 \times 10^{-5}$ for dwarf spheroidal satellite galaxies (dSphs) \cite{Cirelli:2015bda}.

%%%%%%%%%%%%%%%%%%%%%%%%%%%%%%%%%%%%%%%%%%%%%%%%%%
\section{Particle spectra}\label{Sec:Spectra}
%%%%%%%%%%%%%%%%%%%%%%%%%%%%%%%%%%%%%%%%%%%%%%%%%%
Knowing the cross section of dark matter annihilation is insufficient to obtain limits from indirect detection. One must also know the spectra of particles produced in such collisions. We will refer to these spectra as $dN_f/dE$, with $f$ representing the final state of the collision. We will be interested in the flux of gamma rays. The spectra must be computed both for regular annihilation to standard model particles and semi-annihilation.

For regular annihilation, we reuse the spectra of Ref.~\cite{Cirelli:2010xx}. The gamma ray spectra are provided for different 2-to-2 annihilation processes and a series of masses between 5~GeV and 100~TeV. For a generic mass, the spectrum is interpolated from this grid. As this is a rather small effect, we neglect polarization by taking the polarization-averaged spectra of massive gauge bosons. Since it is not available in Ref.~\cite{Cirelli:2010xx}, we have generated ourselves the spectra associated to $ZA$ using \texttt{Pythia}~8.310~\cite{Bierlich:2022pfr}.

For semi-annihilation, we can take the limit of small mixing between $\phi_1$ and $\phi_2$. Particles are produced in two steps. First, the semi-annihilation process itself can take one of the following forms:
\begin{equation}\label{eq:Spectra1}
  \phi_1^i \phi_1^{i'} \to (\phi_2^0)^\dagger h, \quad
  \phi_1^i \phi_1^{i'} \to (\phi_2^0)^\dagger Z, \quad
  \phi_1^i \phi_1^{i'} \to (\phi_2^-)^\dagger W^-,
\end{equation}
where $\phi_1^{i'}$ corresponds to the component of $\phi_1$ with opposite charge to $\phi_1^i$ and we labelled the components of $\phi_2$ in terms of their charge. As mentioned in Sec.~\ref{Sec:SE}, these three processes have cross sections of ratios 1:1:2. Second, the resulting $\phi_2^\dagger$ component will decay through one of three potential channels for each of its possible charges:
\begin{equation}\label{eq:Spectra2}
  \begin{aligned}
    & (\phi_2^0)^\dagger \to (\phi_1^0)^\dagger h, \quad &
    & (\phi_2^0)^\dagger \to (\phi_1^0)^\dagger Z, \quad
    & (\phi_2^0)^\dagger \to (\phi_1^+)^\dagger W^+,\\
    & (\phi_2^-)^\dagger \to (\phi_1^-)^\dagger h, \quad &
    & (\phi_2^-)^\dagger \to (\phi_1^-)^\dagger Z, \quad
    & (\phi_2^-)^\dagger \to (\phi_1^0)^\dagger W^+.
  \end{aligned}
\end{equation}
If the resulting $\phi_1^\dagger$ component is not neutral, it will decay to the neutral component of $\phi_1^\dagger$ and some soft particles that can be safely neglected. The relevant branching ratios are:
\begin{equation}\label{eq:phi2BR}
  \begin{aligned}
    & \text{BR}(\phi_2^0 \to \phi_1^0 h) = \text{BR}(\phi_2^0 \to \phi_1^0 Z) = \frac{n_2}{4n_1},        & & \text{BR}(\phi_2^0 \to \phi_1^+ W^-) = \frac{2n_1 - n_2}{2n_1},\\
    & \text{BR}(\phi_2^- \to \phi_1^- h) = \text{BR}(\phi_2^- \to \phi_1^- Z) = \frac{2n_1 - n_2}{4n_1}, & & \text{BR}(\phi_2^- \to \phi_1^0 W^-) = \frac{n_2}{2n_1}.
  \end{aligned}
\end{equation}
There are therefore nine possible combinations of semi-annihilation and decay processes, resulting in nine different spectra. \texttt{Pythia} is used once more to generate them. They are produced on a two-dimensional grid of $m_1$ and $m_2/m_1$ and interpolated as necessary for a given set of masses. The spectra for the semi-annihilation of antiparticles can safely be approximated as identical.

%%%%%%%%%%%%%%%%%%%%%%%%%%%%%%%%%%%%%%%%%%%%%%%%%%
\section{Indirect detection constraints}\label{Sec:EC}
%%%%%%%%%%%%%%%%%%%%%%%%%%%%%%%%%%%%%%%%%%%%%%%%%%
In this section, we present the different indirect detection constraints that we will consider.

\subsection{HESS inner galaxy survey}\label{sSec:HESS}
The High Energy Stereoscopic System (HESS) is a system of five ground-based Cherenkov telescopes located in Namibia. Their most recent measurements of the gamma ray spectrum emanating from the galactic center are presented in Ref.~\cite{HESS:2022ygk}, which we will use to constrain the dark matter annihilation and semi-annihilation cross sections.

The HESS experiment looks for gamma rays in a series of regions of interest (ROI) referred to as the ON regions. These consist of 25 concentric rings centered around the galactic center, with inner radii ranging from 0.5$^{\circ}$ to 2.9$^{\circ}$ and width of 0.1$^{\circ}$. A set of filters are applied to hide known gamma ray sources. In order to determine if any excess is present, a series of OFF regions are defined symmetrically from the pointing direction, including the same filters. These regions are assumed to contain a lower amount of dark matter but to produce a similar background. A set of fourteen different pointing directions were considered. The original analysis made use of ROI of both finite angular and energy ranges. However, only results for which the energy bins are combined are publicly available. This typically decreases the limits, but not much for the continuum spectra.

The J-factor of the ROI $i$ is defined as
\begin{equation}\label{eq:Jfactor}
  J_i = \int_{\Delta \Omega_i} \int_{\text{l.o.s.}} \rho^2(r(s, \Omega)) ds d\Omega,
\end{equation} 
where $\rho$ is the dark matter mass density, $r = (s^2 + r^2_\odot - 2 r_\odot s \cos \theta)^{1/2}$ is the distance of a given point from the center of the galaxy, $\theta$ is the angle between the direction of observation and the galactic center, $\Delta \Omega_i$ is the angular range of the ROI and l.o.s. stands for line of sight. The HESS experiment assumes the distance between the sun and the galactic center $r_\odot$ to be 8.5~kpc. The J-factors are evaluated for both the ON and OFF regions. For the OFF regions, they are computed for each pointing direction and averaged. 

Since the distribution of dark matter in the galactic center is still not firmly established, we will consider different dark matter density profiles. The Einasto profile is given by
\begin{equation}\label{eq:Einasto}
  \rho_\text{E}(r) = \rho_s \exp\left[-\frac{2}{\alpha_s}\left(\left(\frac{r}{r_s}\right)^{\alpha_s} -1\right)\right]. 
\end{equation}
We will consider two choices of parameters:
\begin{itemize}
  \item Einasto 1: $\rho_s = 0.079 \text{ GeV/cm}^3$, $r_s = 20.0$~kpc and $\alpha_s = 0.17$,
  \item Einasto 2: $\rho_s = 0.033 \text{ GeV/cm}^3$, $r_s = 28.4$~kpc and $\alpha_s = 0.17$.
\end{itemize}
We will also consider the Navarro-Frenk-White (NFW) profile given by
\begin{equation}\label{eq:NFW}
  \rho_\text{NFW}(r) = \rho_s \left(\frac{r}{r_s}\left(1 + \frac{r}{r_s}\right)^2\right)^{-1},
\end{equation}
where we will use $\rho_s = 0.307 \text{ GeV/cm}^3$ and $r_s = 21.0$~kpc. The distributions above correspond to the choices of Ref.~\cite{HESS:2022ygk} and have been used to validate our results. Furthermore, we will consider the Einasto distributions with flat cores. For this, we use the Einasto 1 profile and assume the density to be constant below a core radius $r_c \in \left\{1, 2, 4\right\}$~kpc. Constraints are expected to decrease when the core size increases, as the distinction between the number of signals in the ON and OFF regions shrinks and as there is simply less dark matter.

The number of signals in the ROI $i$ is given by
\begin{equation}\label{eq:HESSSignal}
  N^s_i = \sum_f \frac{\langle \sigma v \rangle_f J_i}{16 \pi m_1^2} T_{\text{obs}} \int_{E_{\text{th}}}^{m_1} \int_0^\infty \frac{dN_f}{dE}(E) R(E, E') A_\text{eff}(E) dE dE'.
\end{equation}
The observation time $T_{\text{obs}}$ corresponds to 546 hours. The threshold energy $E_{\text{th}}$ is 200~GeV. The resolution $R(E, E')$ is taken as a Gaussian of $\sigma/E = 10\%$ and the acceptance $ A_\text{eff}(E)$ is read from Ref.~\cite{HESS:2022ygk}. Eq.~\eqref{eq:HESSSignal} takes into account that the dark matter candidate is distinct from its antiparticle, which introduces an additional factor of 2 in the denominator.

The likelihood is given by
\begin{equation}\label{eq:HESSlikelihood}
  \mathcal{L}(\mathbf{N}^s, \mathbf{N}^b) = \prod_i
                       \frac{(N^s_i + N^b_i)^{N_{\text{ON}, i}}}{N_{\text{ON}, i}!} e^{-(N^s_i + N^b_i)}
                       \frac{(N^s_i{}' + N^b_i)^{N_{\text{OFF}, i}}}{N_{\text{OFF}, i}!} e^{-(N^s_i{}' + N^b_i)}.
\end{equation}
The number of signals in the ON and OFF regions are respectively labelled as $N^s_i$ and $N^s_i{}'$. The number of backgrounds $N^b_i$ are treated as nuisance parameters. Systematic uncertainties are neglected, but are expected to have an impact of less than $\mathcal{O}(20\%)$ on the cross section limits \cite{HESS:2022ygk}. The quantities $N_{\text{ON}, i}$ and $N_{\text{OFF}, i}$ correspond to the number of observed events in the ON and OFF regions respectively and are available in Ref.~\cite{HESS:2022ygk}.

A profile likelihood is used as a test statistic and defined as \cite{Cowan:2010js}:
\begin{equation}\label{eq:HESSTS}
  \text{TS} = \begin{cases}
  -2\ln \frac{\mathcal{L}(\mathbf{N}^s(\langle \sigma v \rangle),\; \doublehat{\mathbf{N}}{}^b(\langle \sigma v \rangle)}{\mathcal{L}(0,\; \doublehat{\mathbf{N}}{}^b(0)}, 
  & \text{if $\widehat{\langle \sigma v \rangle} < 0$,}\\
  -2\ln \frac{\mathcal{L}(\mathbf{N}^s(\langle \sigma v \rangle),\; \doublehat{\mathbf{N}}{}^b(\langle \sigma v \rangle)}{\mathcal{L}(N^s(\widehat{\langle \sigma v \rangle}),\; \widehat{\mathbf{N}}^b(\widehat{\langle \sigma v \rangle})}, 
  & \text{if $0 \leq \widehat{\langle \sigma v \rangle} \leq \langle \sigma v \rangle$,}\\
  0, & \text{otherwise}.
  \end{cases} 
\end{equation}
We have defined $\langle \sigma v \rangle = \sum_f \langle \sigma v \rangle_f$. Quantities with two hats maximize the likelihood for a fixed $\langle \sigma v \rangle$. Quantities with one hat maximize the likelihood when $\langle \sigma v \rangle$ is allowed to vary. Following Ref.~\cite{HESS:2022ygk}, a point is considered excluded at 95\% one-sided confidence level if $\text{TS} > 2.71$. We have validated our results by verifying that we reproduce their limits on regular dark matter annihilation to standard model particles.

\subsection{Fermi-LAT dwarf spheroidal satellite galaxies survey}\label{sSec:FermiLAT}
Fermi-LAT is an imaging gamma-ray detector aboard the Fermi Gamma-ray Space Telescope. Ref.~\cite{Fermi-LAT:2015att} provides their limits on dark matter annihilation rate coming from the observation of 15 dwarf spheroidal satellite galaxies over a period of 6 years using the \texttt{Pass} 8 event-level analysis.

A series of energy bins ranging from 0.5~GeV to 500~GeV are defined. The energy flow from a dSph $i$ in an energy bin $j$ is given by
\begin{equation}\label{eq:Eflow}
  \Phi^E_{ij} = \sum_f \frac{\langle \sigma v \rangle_f J_i}{16 \pi m_1^2} \int_{E^\text{min}_j}^{E^\text{max}_j} E \frac{dN_f}{dE}(E) dE,
\end{equation}
where $E^\text{min}_j$ and $E^\text{max}_j$ represent respectively the lower and upper limits of the energy bin. The likelihood of different values of $\Phi^E_{ij}$ are available in Ref.~\cite{Fermi-LAT:2015att} and are referred to as $\mathcal{L}_{ij}$. One can then define a likelihood for a dSph $i$ as
\begin{equation}\label{eq:FermiLATLikelihood1}
  \mathcal{L}_i(\langle \sigma v \rangle,\; J_i) = \prod_j \mathcal{L}_{ij}(\langle \sigma v \rangle,\; J_i).
\end{equation}
The J-factors $J_i$ are treated as nuisance parameters with an associated likelihood given by a log-normal distribution
\begin{equation}\label{eq:FermiLATLikelihood2}
  \mathcal{L}_{J_i}(J_i) = \frac{1}{\ln(10)J_{\text{obs}, i}\sqrt{2\pi} \sigma_i} e^{-\frac{(\log_{10} (J_i) - \log_{10} (J_{\text{obs}, i}))^2}{2\sigma_i^2}},
\end{equation}
where the observed J-factor $J_{\text{obs}, i}$ and the uncertainties $\sigma_i$ on $\log_{10}(J_{\text{obs}, i})$ are available in Ref.~\cite{Fermi-LAT:2015att}. The total likelihood is then defined as
\begin{equation}\label{eq:FermiLATLikelihood3}
  \mathcal{L}(\langle \sigma v \rangle,\; \mathbf{J}) = \prod_i \mathcal{L}_i(\langle \sigma v \rangle,\; J_i) \times \mathcal{L}_{J_i}(J_i).
\end{equation}
Systematic uncertainties are neglected. The test statistic is chosen as
\begin{equation}\label{eq:FermiLATTS}
  \text{TS} = - 2 \ln \frac{\mathcal{L}(\langle \sigma v \rangle,\; \doublehat{\mathbf{J}})}{\mathcal{L}(\widehat{\langle \sigma v \rangle},\; \widehat{\mathbf{J}})}.
\end{equation}
Quantities with a double hat maximize the likelihood for a given $\langle \sigma v \rangle$. Quantities with a single hat maximize the likelihood when $\langle \sigma v \rangle$ is allowed to vary. Following the conventions of Ref.~\cite{Fermi-LAT:2015att}, a point is considered excluded at 95\% confidence level if $\text{TS} > 2.71$. We verified that we could reproduce their limits on regular dark matter annihilation with good accuracy. Typically, limits from Fermi-LAT will be subdominant to HESS if the profile density is cuspy, but can be stronger if the core is large.

\subsection{Gamma ray line searches in the galactic center}\label{sSec:Lines}
Because the Sommerfeld effect can result in the annihilation of two charged particles, a line signal is expected in the gamma ray spectra around $m_1$. We will therefore also consider constraints from line searches, although they are ill-suited for semi-annihilation and will turn out to be subdominant.

The MAGIC experiment consists of two imaging atmospheric Cherenkov telescopes located in Spain. Ref.~\cite{MAGIC:2022acl} contains their latest gamma ray line search in the galactic center. Unfortunately, only their limits on the annihilation to two photons cross section are publicly available. We will therefore use this constraint exclusively, though taking into account the fact that the dark matter candidate is distinct from its antiparticle in our case. The limits of Ref.~\cite{MAGIC:2022acl} are presented for four dark matter density profiles. 
\begin{itemize}
  \item Einasto 1 as defined in Sec.~\ref{sSec:HESS}.
  \item NFW with $\rho_s = 0.0768 \text{ GeV/cm}^3$, $r_s = 21.0$~kpc and $r_\odot = 8.5$~kpc.
  \item Hernquist-Zhao:
  \begin{equation}\label{eq:HZ}
    \rho_{\text{HZ}}(r) = \frac{2^{\frac{\beta - \gamma}{\alpha}}\rho_s}{\left(\frac{r}{r_s}\right)^\gamma\left[1 + \left(\frac{r}{r_s}\right)^\alpha \right]^\frac{\beta - \alpha}{\gamma}},
  \end{equation}
  with $\alpha = 1$, $\beta = 3$, $\gamma = 0$, $\rho_s = 0.431\text{ GeV/cm}^3$, $r_s = 7.7$~kpc and $r_\odot = 8.21$~kpc.
  \item Burkert:
  \begin{equation}\label{eq:Burkert}
    \rho_{\text{Burkert}}(r) = \frac{\rho_s}{\left(1 + \frac{r}{r_s}\right)\left(1 + \frac{r^2}{r_s^2}\right)}
  \end{equation}
  with $\rho_s = 1.568\text{ GeV/cm}^3$, $r_s = 9.26$~kpc, and $r_\odot = 7.94$~kpc.
\end{itemize}
Limits are applied on the sum of the cross section to $AA$ and half the cross section to $ZA$.

The HESS experiment also performed a line search in Ref.~\cite{HESS:2018cbt}. Once again, only their limits on the dark matter annihilation to two photons are publicly available. Their results are presented for the Einasto 1 and NFW dark matter density profiles of Sec.~\ref{sSec:HESS}. Ref.~\cite{Montanari:2023bzn} contains more recent and stronger limits from the HESS experiment. As the results are only preliminary and only presented for the Einasto profile, we will omit this search. The inclusion of these results would not change any qualitative features.

%%%%%%%%%%%%%%%%%%%%%%%%%%%%%%%%%%%%%%%%%%%%%%%%%%
\section{Results}\label{Sec:Results}
%%%%%%%%%%%%%%%%%%%%%%%%%%%%%%%%%%%%%%%%%%%%%%%%%%
We present in this section examples of the constraints from indirect detection on the semi-annihilation of inert scalar multiplets.

\begin{figure}[t!]
    \centering
    \begin{subfigure}{0.49\textwidth}
        \centering
        \caption{}
        \includegraphics[width=\textwidth]{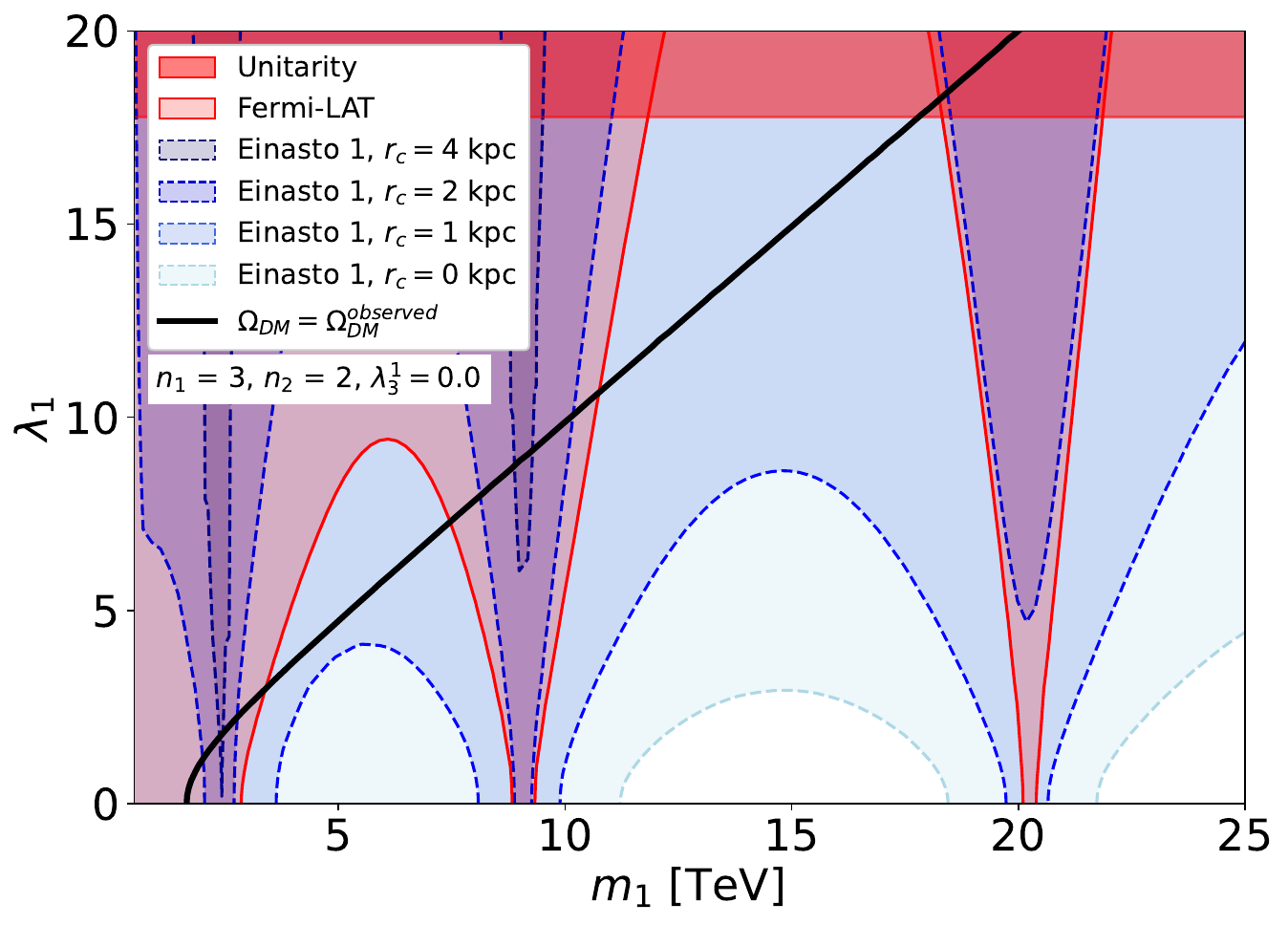}
        \label{fig:Scan1_32}
    \end{subfigure}
    \begin{subfigure}{0.49\textwidth}
        \centering
        \caption{}
        \includegraphics[width=\textwidth]{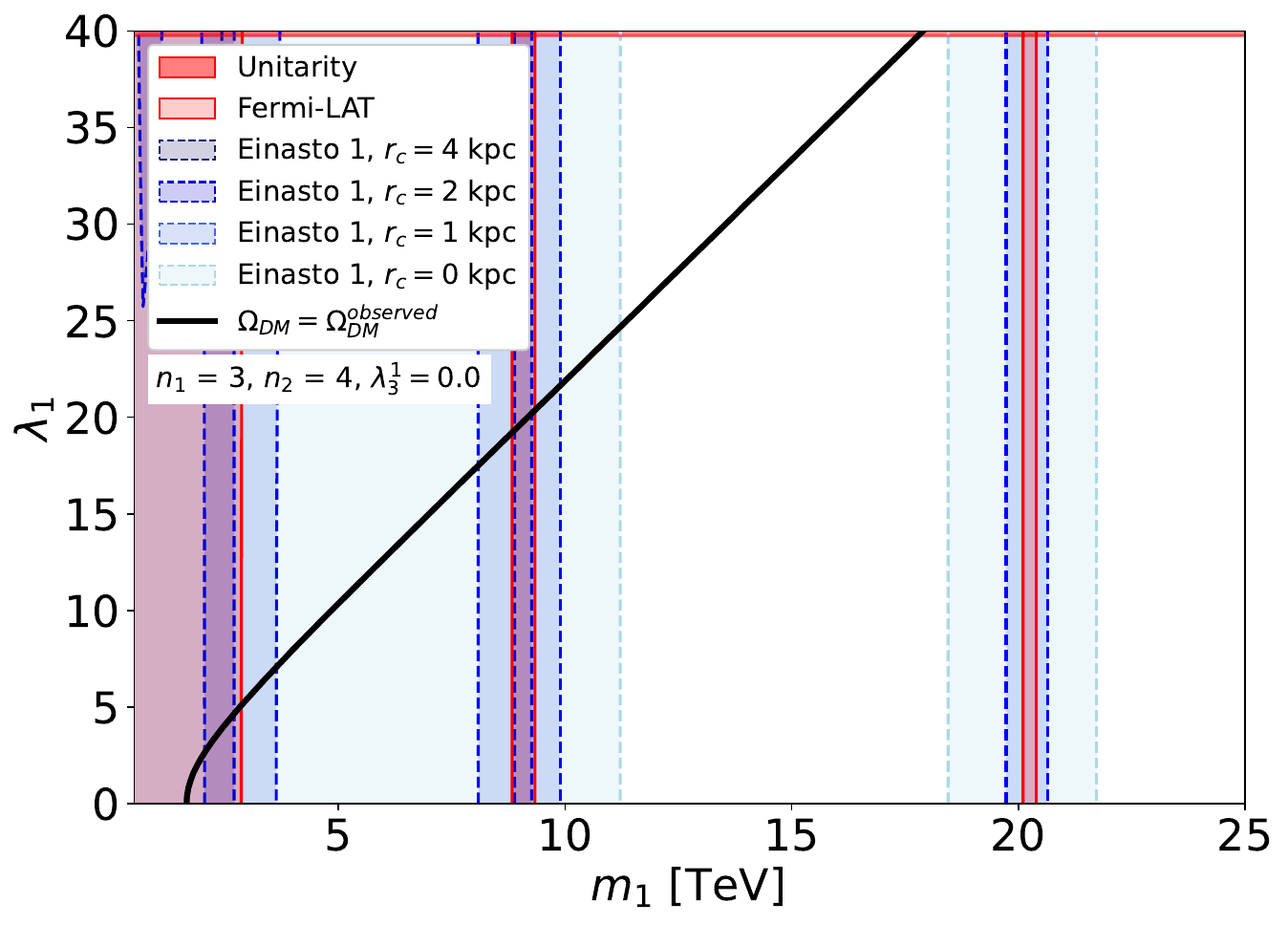}
        \label{fig:Scan1_34}
    \end{subfigure}
    \begin{subfigure}{0.49\textwidth}
        \centering
        \caption{}
        \includegraphics[width=\textwidth]{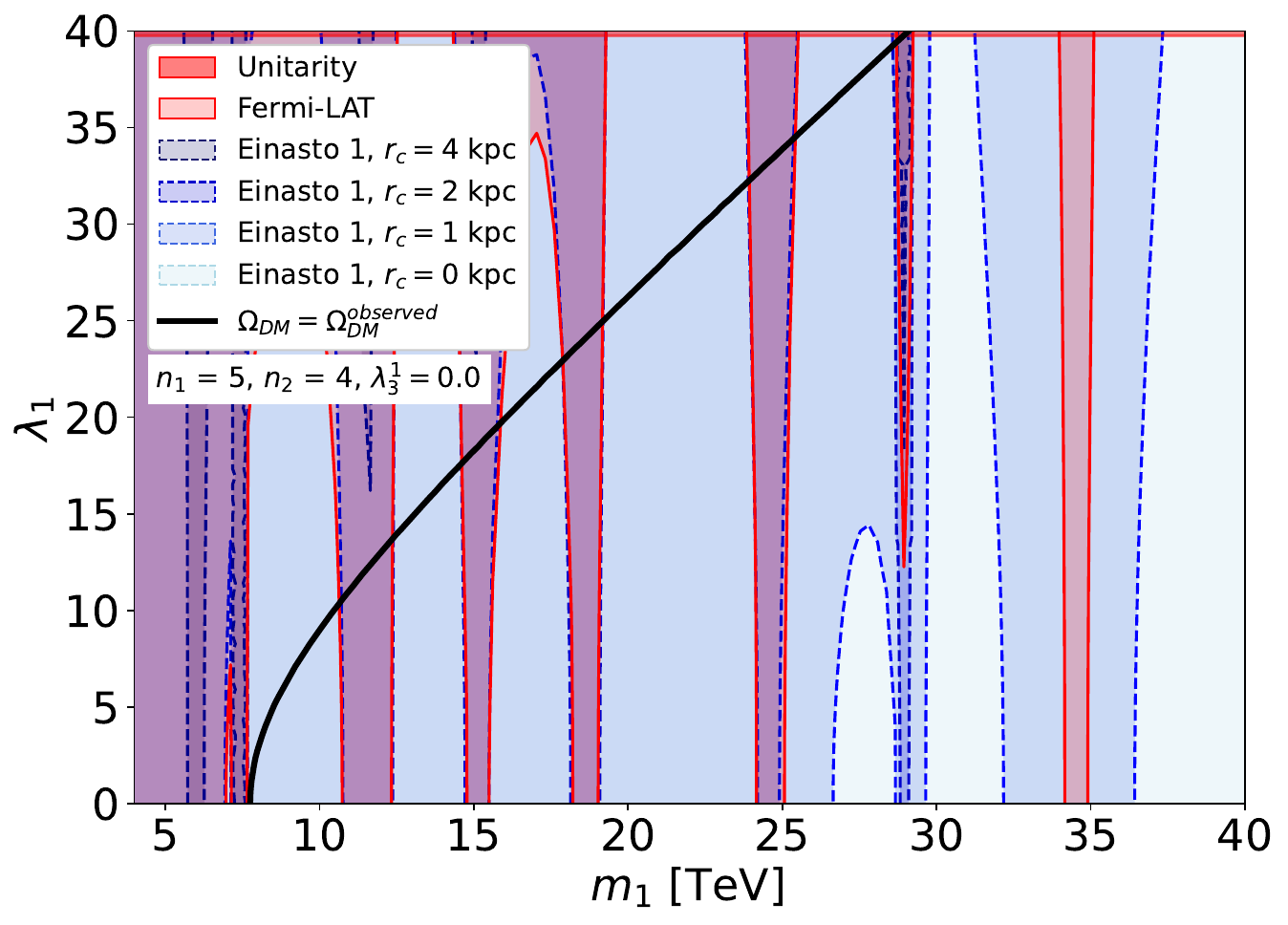}
        \label{fig:Scan1_54}
    \end{subfigure}
    \begin{subfigure}{0.49\textwidth}
        \centering
        \caption{}
        \includegraphics[width=\textwidth]{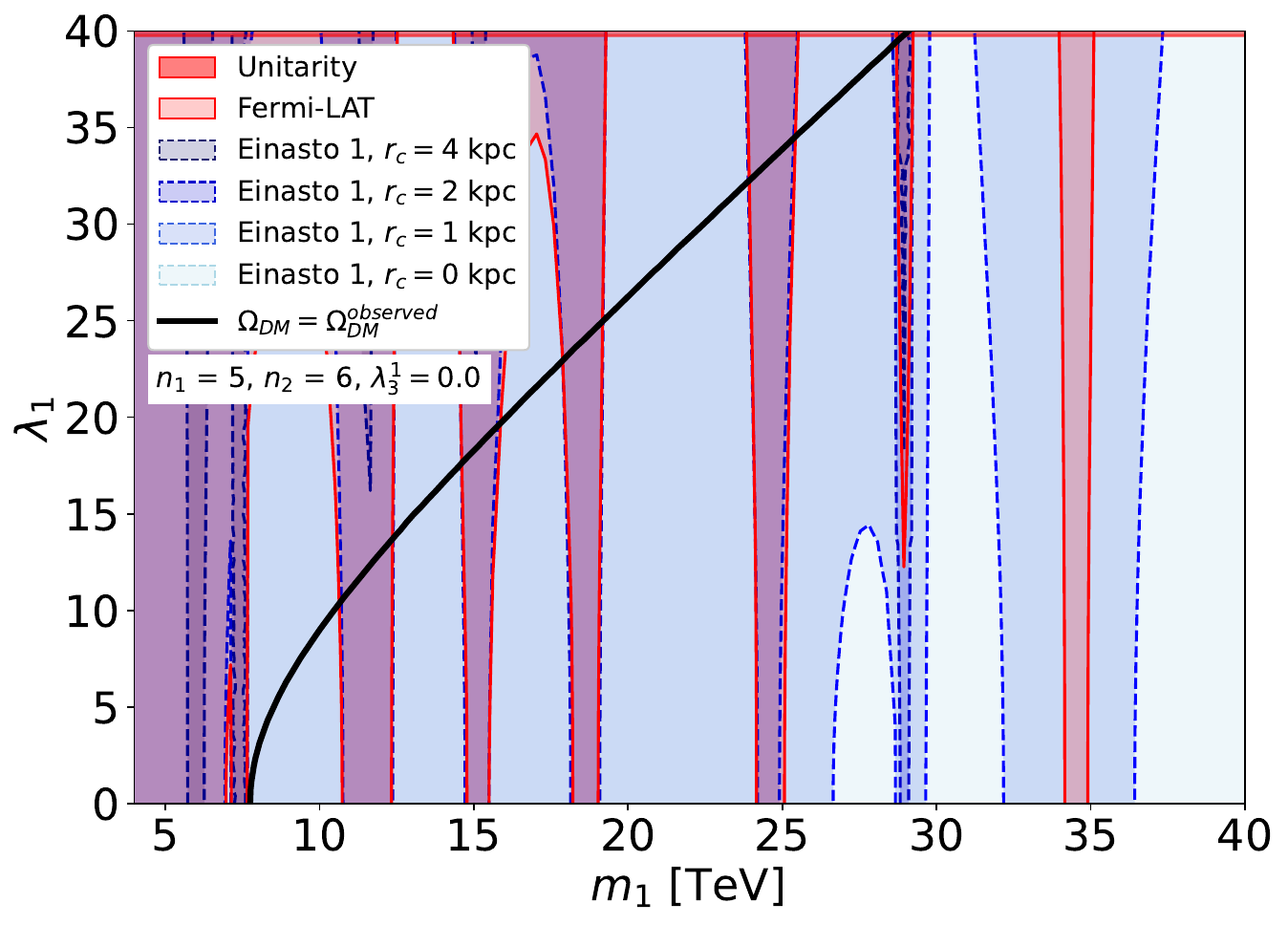}
        \label{fig:Scan1_56}
    \end{subfigure}
    \begin{subfigure}{0.49\textwidth}
        \centering
        \caption{}
        \includegraphics[width=\textwidth]{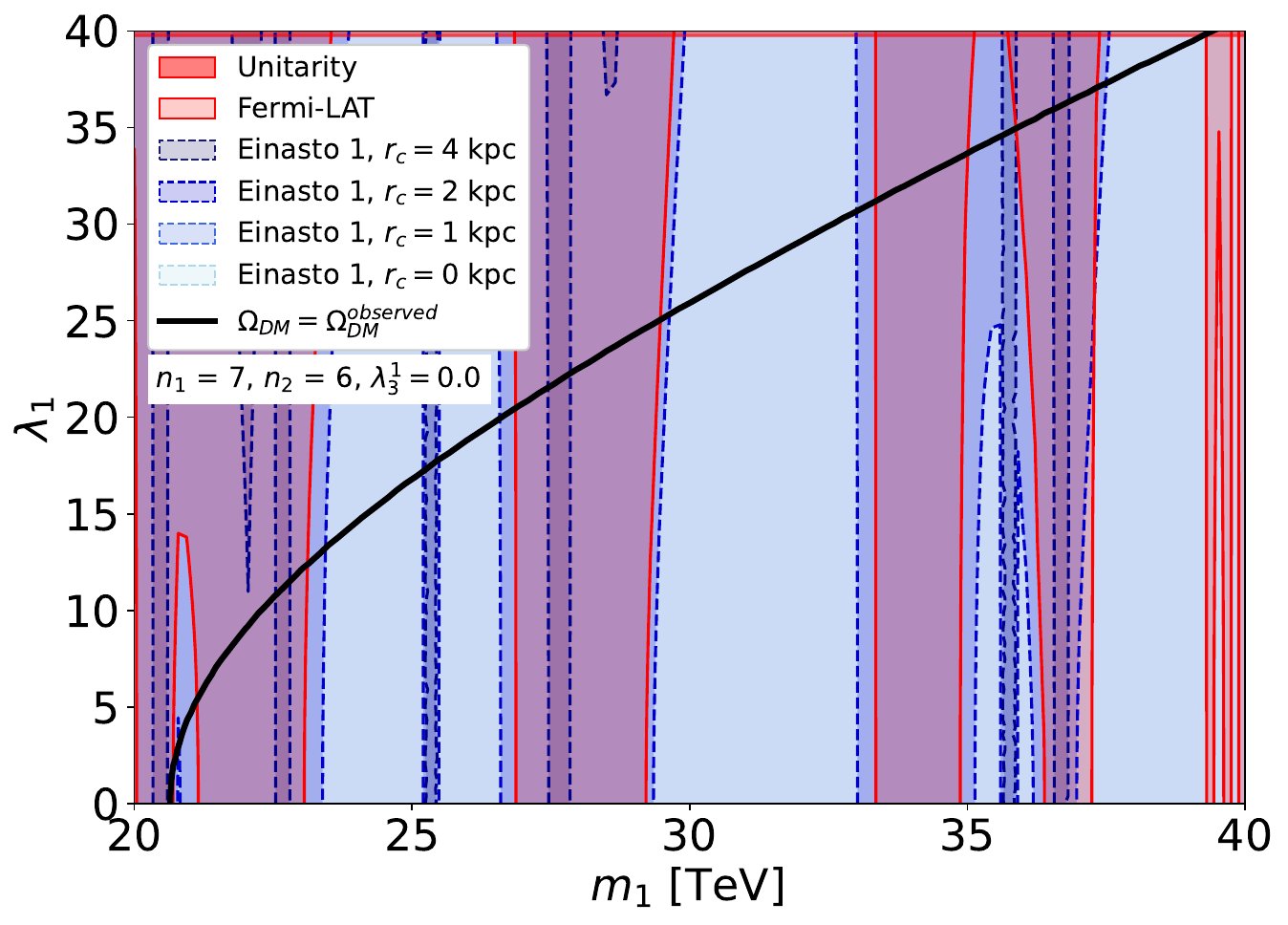}
        \label{fig:Scan1_76}
    \end{subfigure}
    \caption{Constraints from unitarity, Fermi-LAT and the HESS continuum search for the Einasto 1 profiles with different core sizes. The ratio $m_2/m_1 = 1.2$ and all other parameters are zero. The black line matches the observed dark matter abundance.}
    \label{fig:Scan1}
\end{figure}

\begin{figure}[t!]
    \centering
    \begin{subfigure}{0.49\textwidth}
        \centering
        \caption{}
        \includegraphics[width=\textwidth]{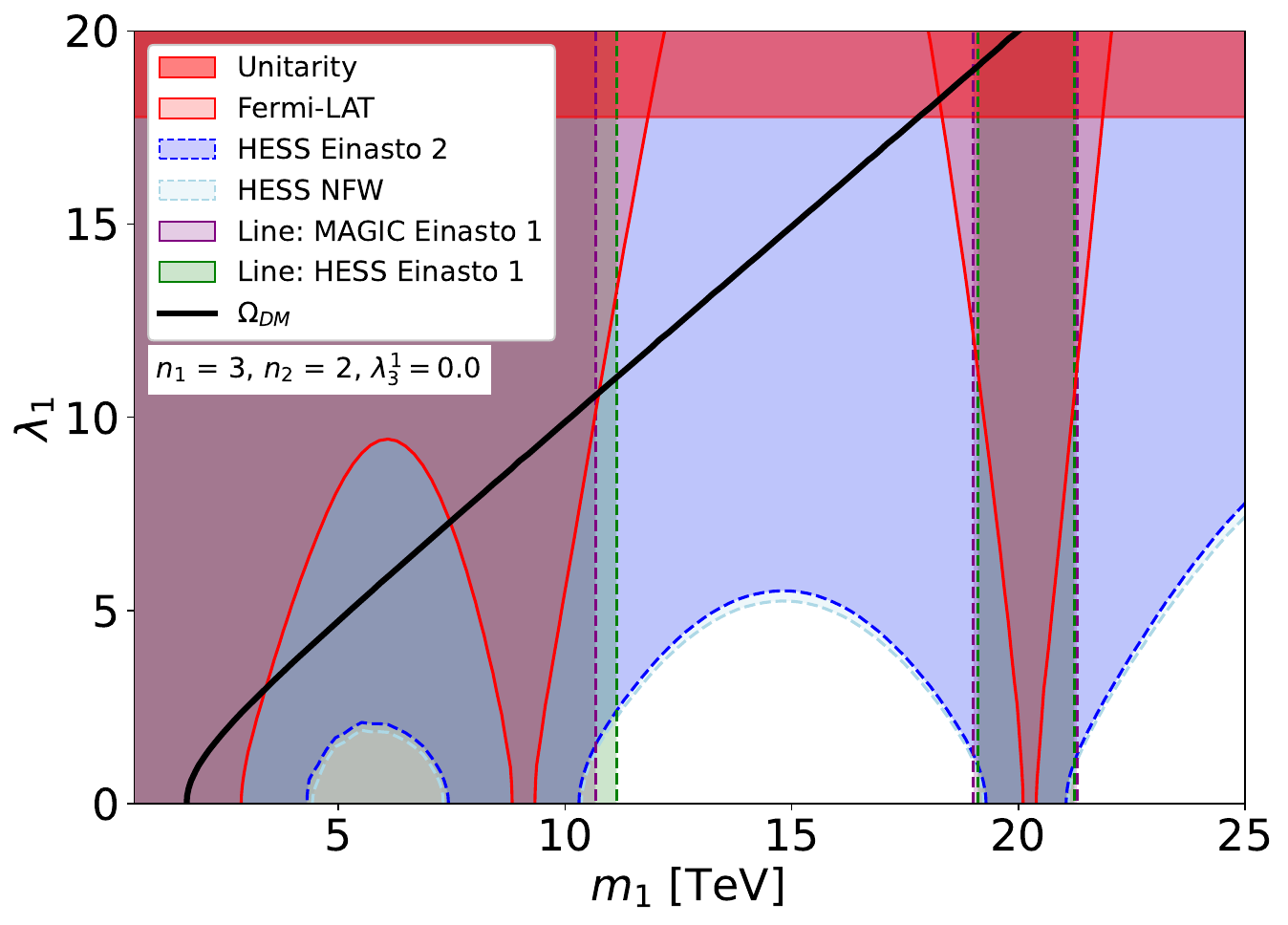}
        \label{fig:Scan2_32}
    \end{subfigure}
    \begin{subfigure}{0.49\textwidth}
        \centering
        \caption{}
        \includegraphics[width=\textwidth]{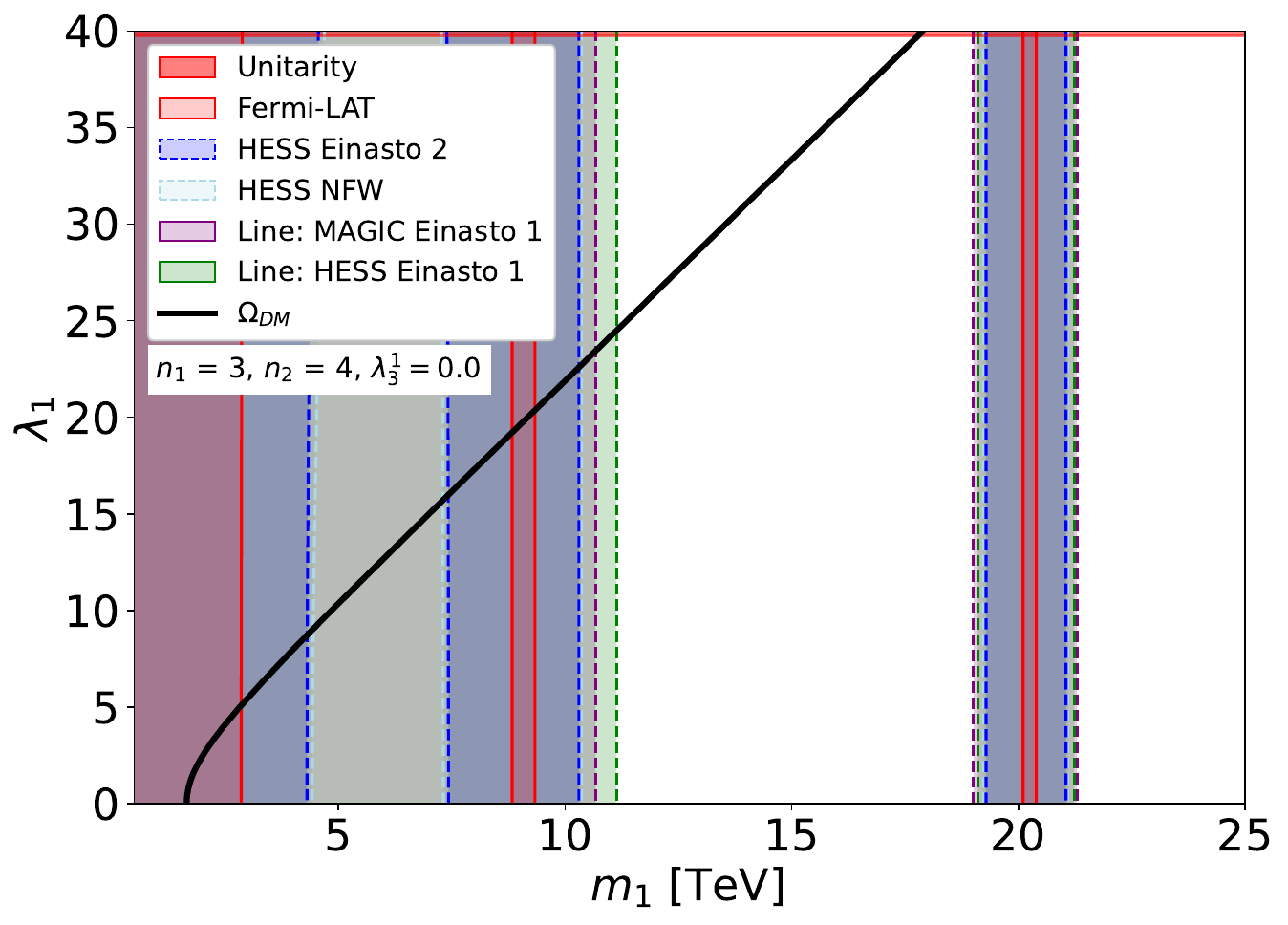}
        \label{fig:Scan2_34}
    \end{subfigure}
    \begin{subfigure}{0.49\textwidth}
        \centering
        \caption{}
        \includegraphics[width=\textwidth]{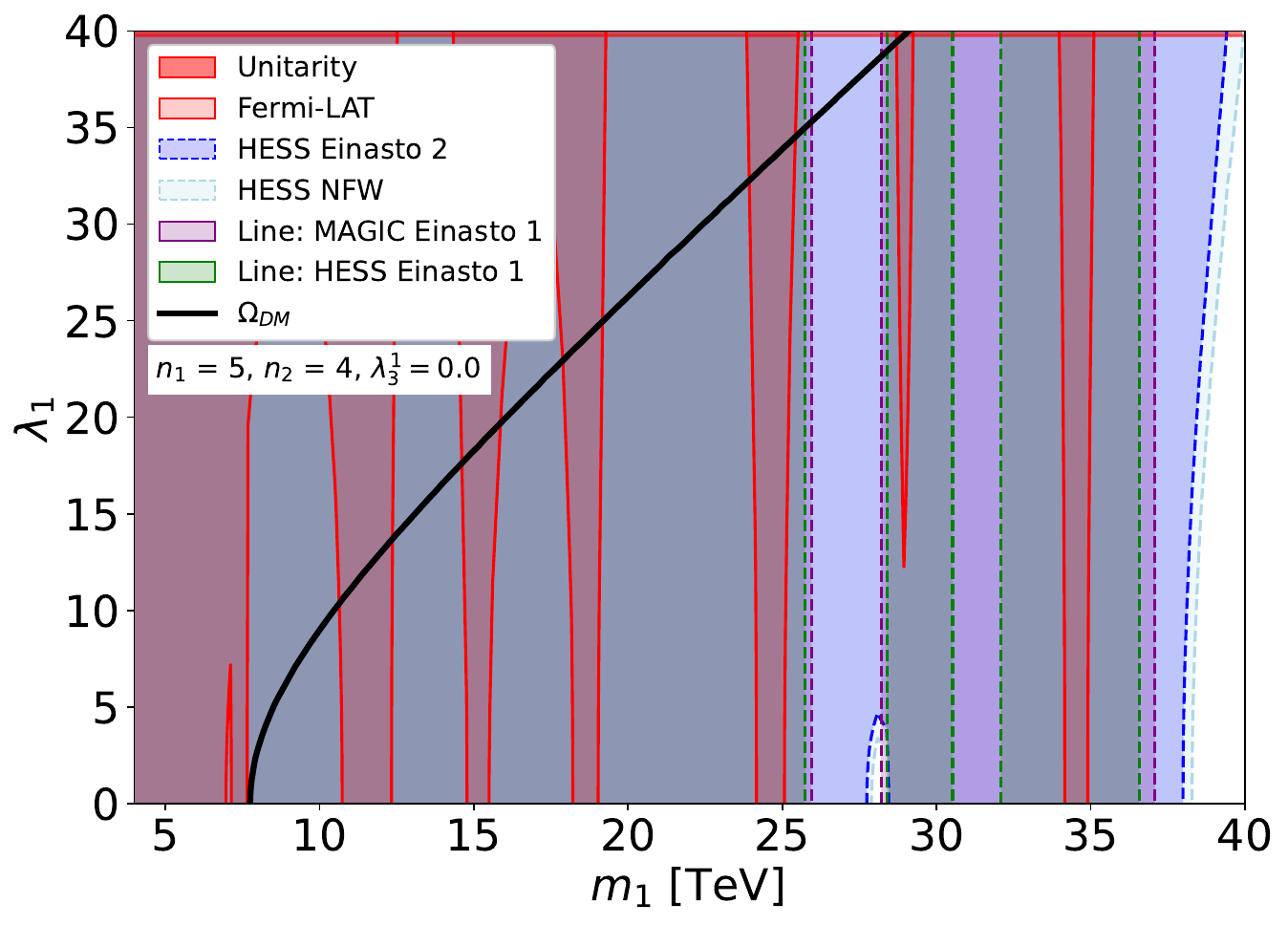}
        \label{fig:Scan2_54}
    \end{subfigure}
    \begin{subfigure}{0.49\textwidth}
        \centering
        \caption{}
        \includegraphics[width=\textwidth]{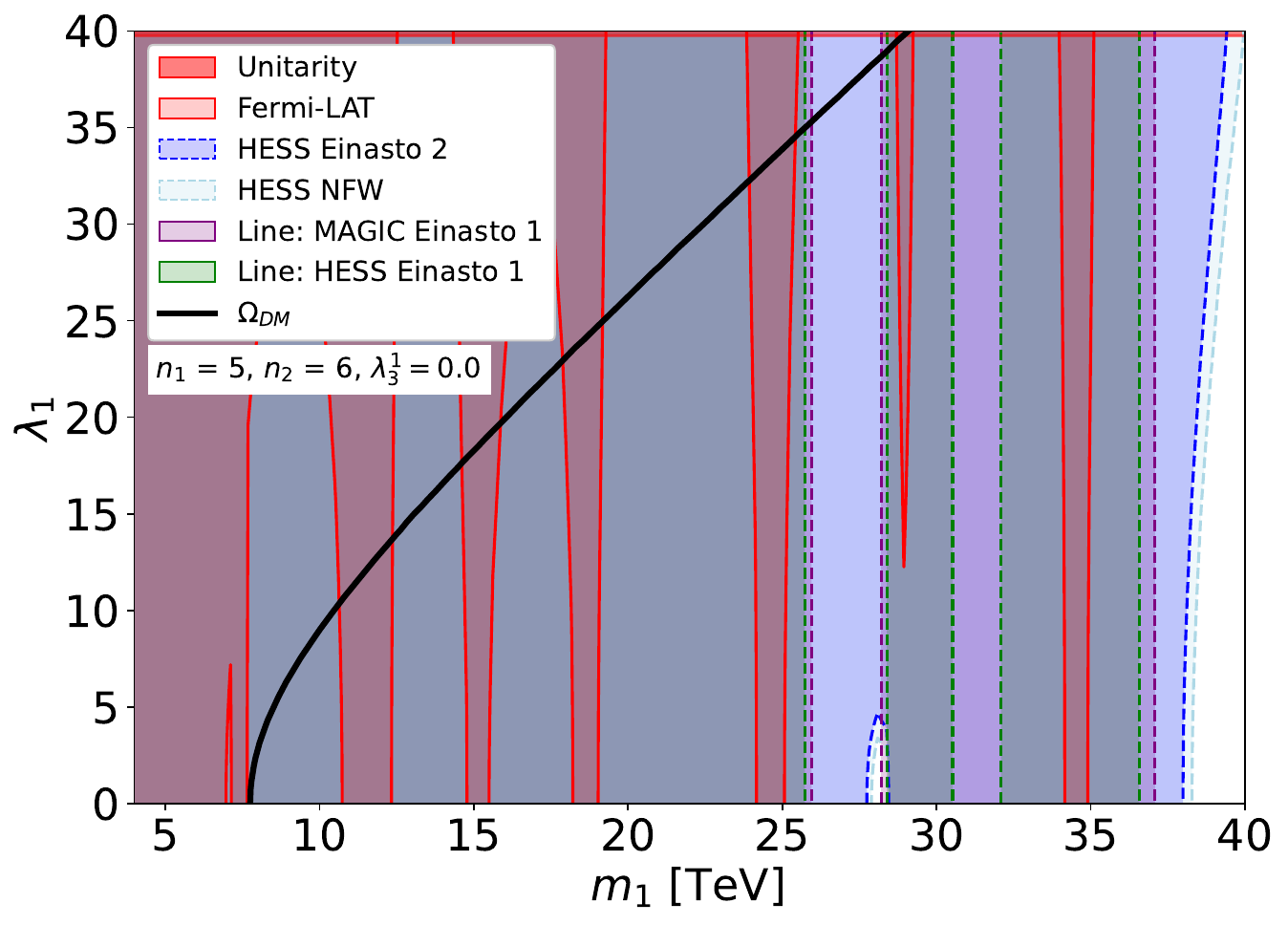}
        \label{fig:Scan2_56}
    \end{subfigure}
    \begin{subfigure}{0.49\textwidth}
        \centering
        \caption{}
        \includegraphics[width=\textwidth]{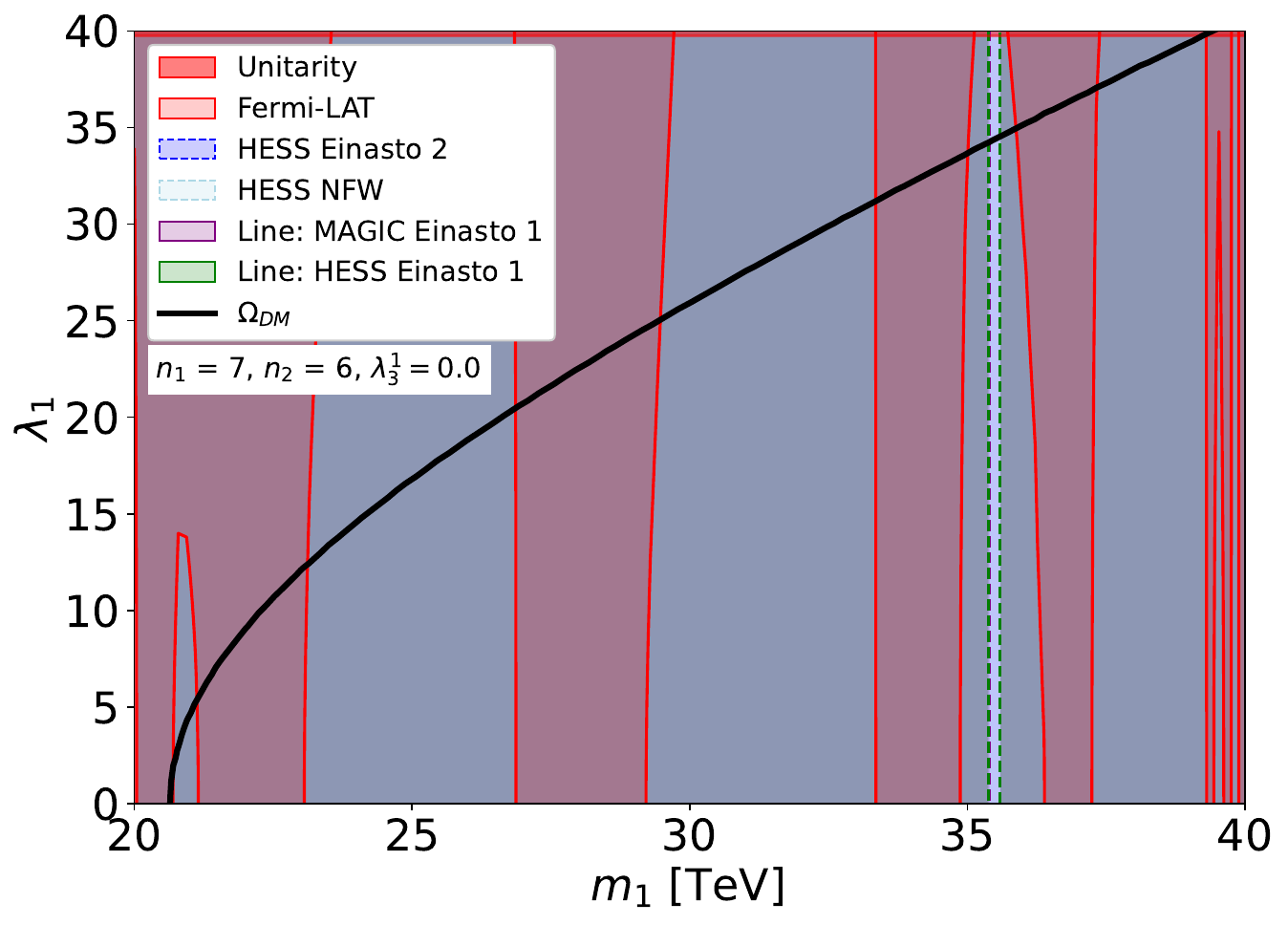}
        \label{fig:Scan2_76}
    \end{subfigure}
    \caption{Constraints from unitarity, Fermi-LAT, the alternative HESS profiles and the line searches. The ratio $m_2/m_1 = 1.2$ and all other parameters are zero. The black line matches the observed dark matter abundance.}
    \label{fig:Scan2}
\end{figure}

Fig.~\ref{fig:Scan1} illustrates the simple case in which only the parameters $m_1$, $m_2$ and $\lambda_1$ are assumed not to be negligible. Results are presented for different choices of multiplet sizes. The ratio $m_2/m_1$ is set to 1.2. This value allows for the semi-annihilation to have a large impact on the dark matter abundance, though any value above 1 and not too close to 2 would lead to qualitatively similar results~\cite{Beauchesne:2024vbo}. It is assumed that $\lambda_5$ has a small non-zero value sufficient for $\phi_1$ and $\phi_2$ to maintain relative chemical equilibrium in the early universe, but not large enough to otherwise impact the abundance \cite{Beauchesne:2024vbo}. The dark red region at the top is excluded by perturbative unitarity, which requires~\cite{Beauchesne:2024vbo}
\begin{equation}\label{eq:UnitarityFinal}
  |\lambda_1| \leq 4 \pi \sqrt{2 R},
\end{equation}
where $R = 4\lfloor n_2/4 \rfloor + 1$. The light red region is excluded by Fermi-LAT. The different shades of blue correspond to the regions excluded by the HESS continuum search for the Einasto 1 profile and cored Einasto 1 with different radii. The thick black line corresponds to the value of $\lambda_1$ that reproduces the observed dark matter abundance under standard thermal freeze-out and was computed using the results of Ref.~\cite{Beauchesne:2024vbo}. Fig.~\ref{fig:Scan2} shows the line constraints and limits for other HESS dark matter density profiles. The line constraints correspond to the strongest by HESS and MAGIC. These constraints are generally subdominant to the HESS Einasto 1 profile with $r_c = 0$ of the continuum search.

For all cases, there exist regions that can explain the correct dark matter abundance and be compatible with Fermi-LAT constraints. For cases other than $n_1 =3$ and $n_2 = 4$, the correct abundance can be compatible with HESS constraints if the galactic core is sufficiently large. This requirement is somewhat more demanding for larger multiplets, as the ability of semi-annihilation to modify the dark matter abundance is relatively more limited for them. The case of $n_1 =5$ and $n_2 = 4$ and $n_1 =5$ and $n_2 = 6$ are almost identical, but can differ more noticeably if $m_2/m_1$ is very close to 1. For $n_1 =3$ and $n_2 = 4$, the limits are almost purely vertical lines, as one would expect from the fact that semi-annihilation processes are Sommerfeld-suppressed in this case. The only exception is for very low mass, where the Sommerfeld effect begins to be less efficient. For these gauge numbers, the correct dark matter abundance can be obtained thermally even in the absence of a core, though this would require a value of $\lambda_1$ close to its unitarity limit.

\begin{figure}[t!]
    \centering
    \begin{subfigure}{0.49\textwidth}
        \centering
        \caption{}
        \includegraphics[width=\textwidth]{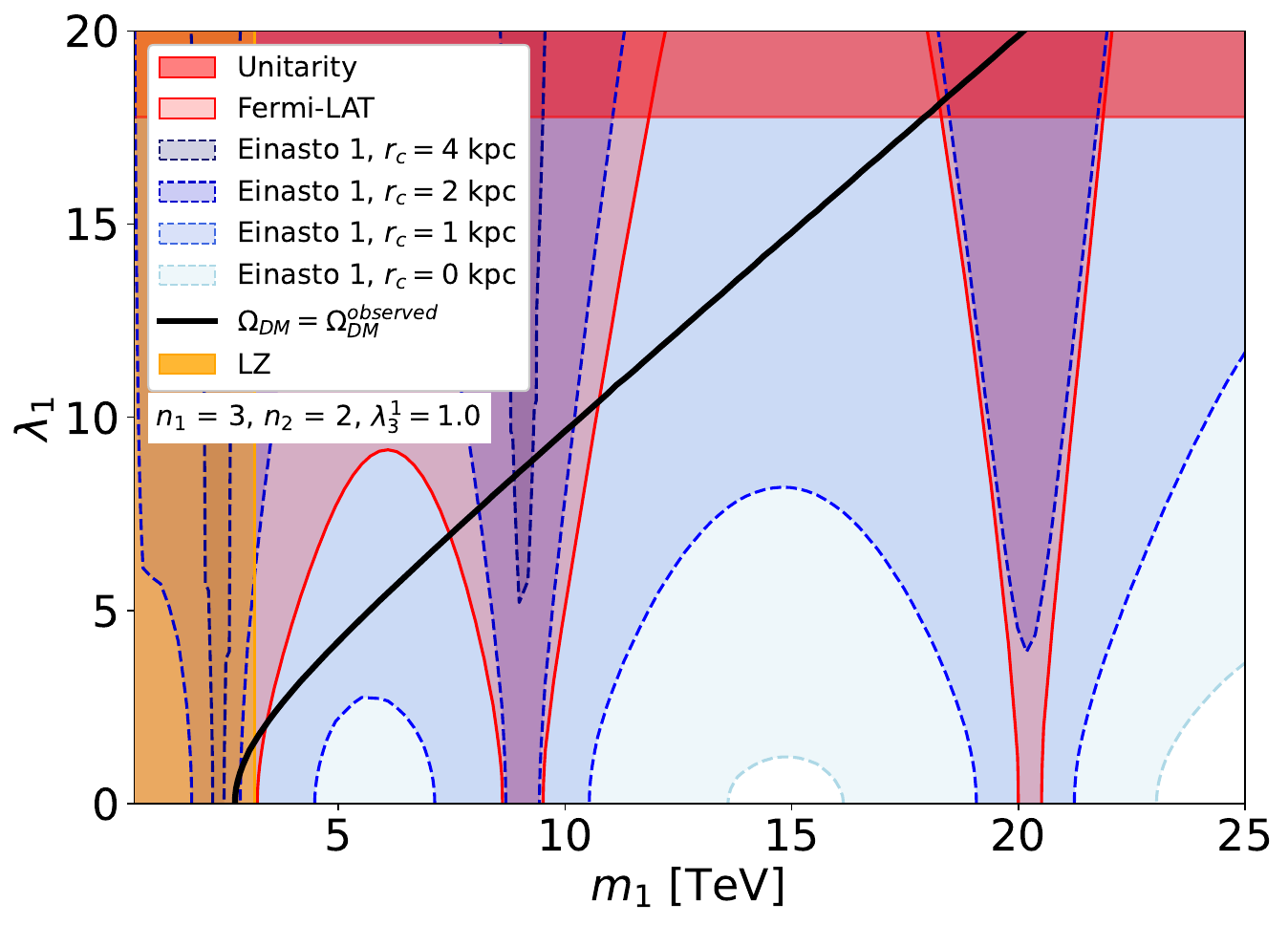}
        \label{fig:Scan3_32a}
    \end{subfigure}
    \begin{subfigure}{0.49\textwidth}
        \centering
        \caption{}
        \includegraphics[width=\textwidth]{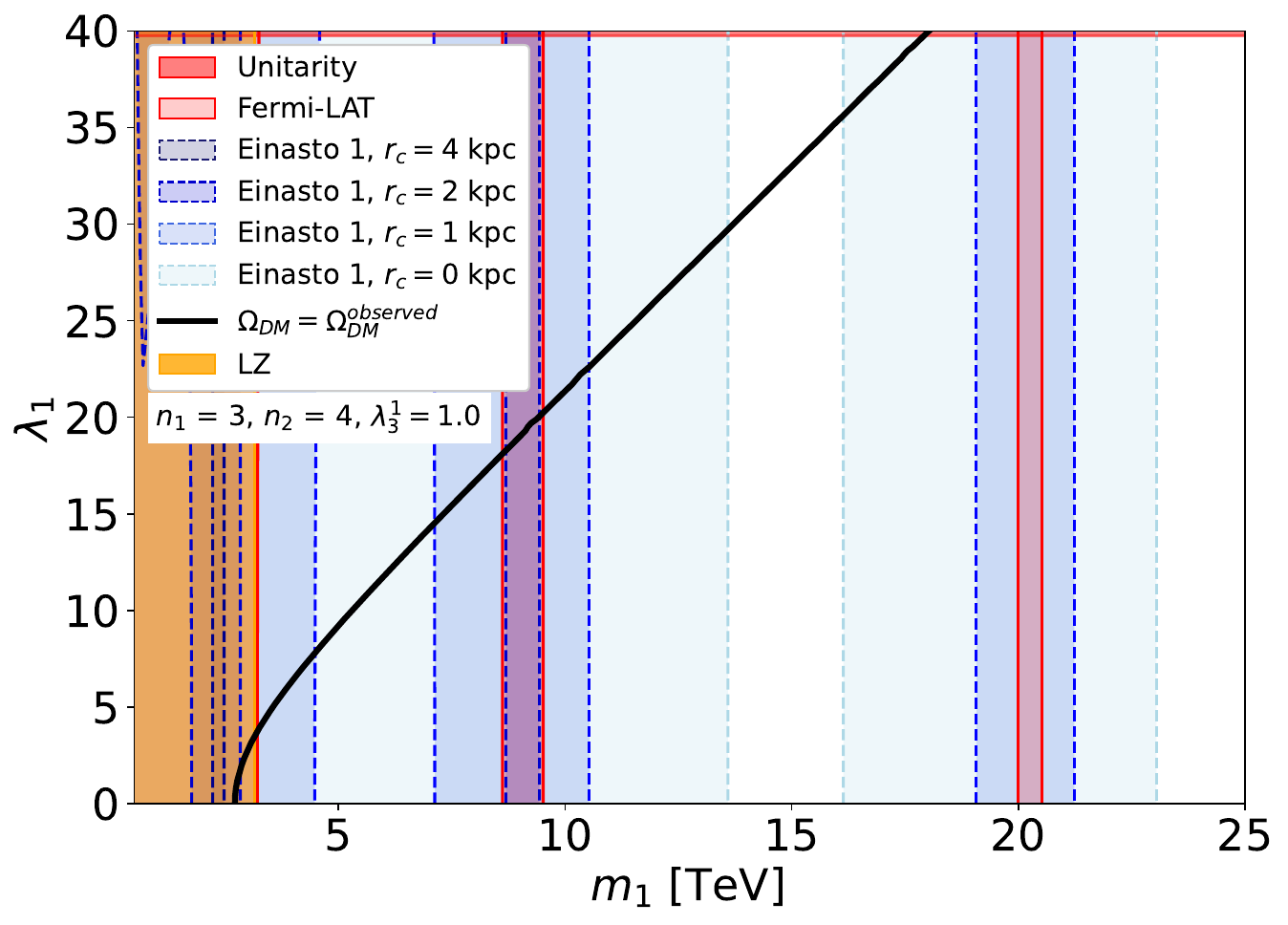}
        \label{fig:Scan3_34a}
    \end{subfigure}
    \begin{subfigure}{0.49\textwidth}
        \centering
        \caption{}
        \includegraphics[width=\textwidth]{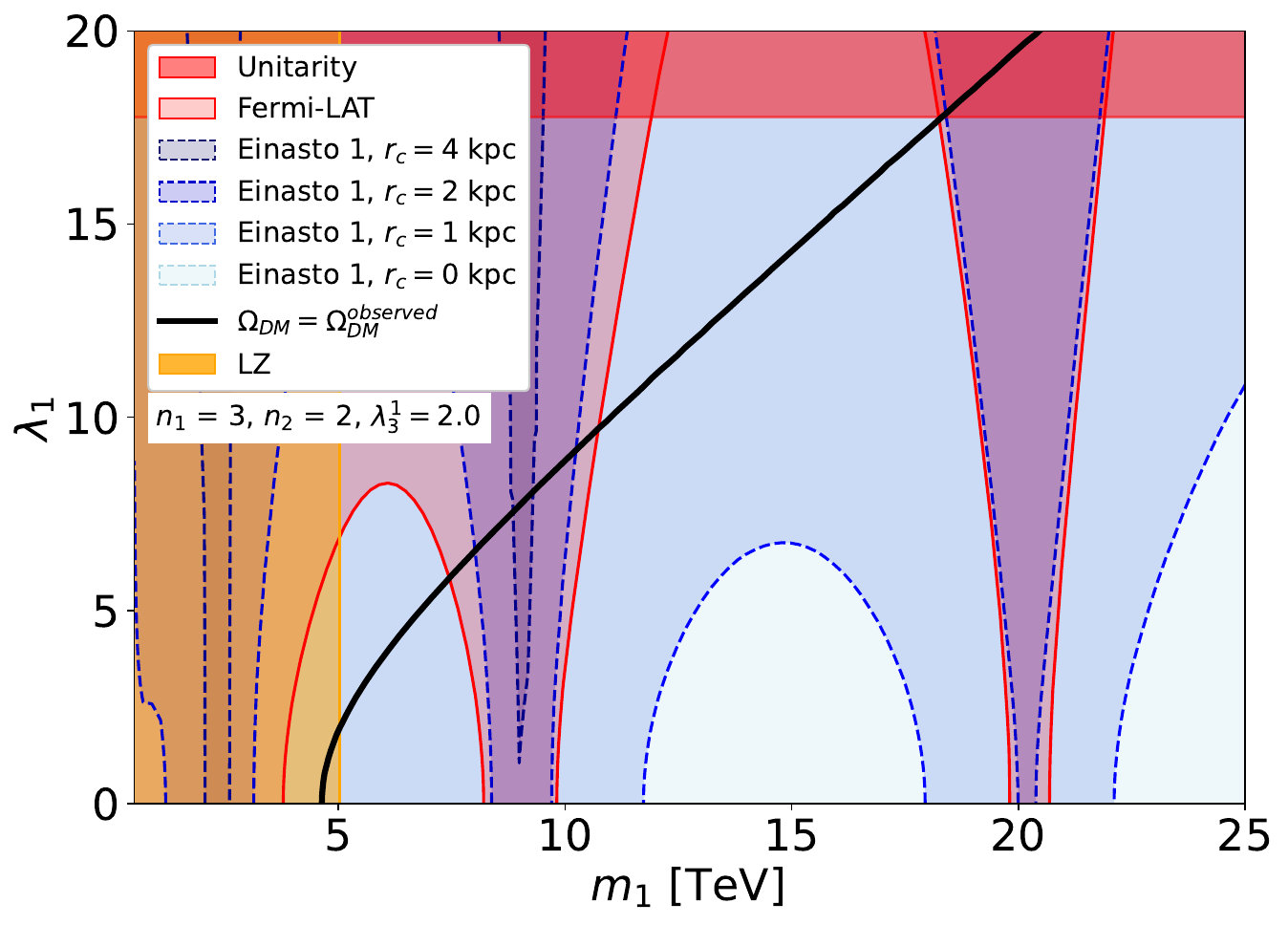}
        \label{fig:Scan3_32b}
    \end{subfigure}
    \begin{subfigure}{0.49\textwidth}
        \centering
        \caption{}
        \includegraphics[width=\textwidth]{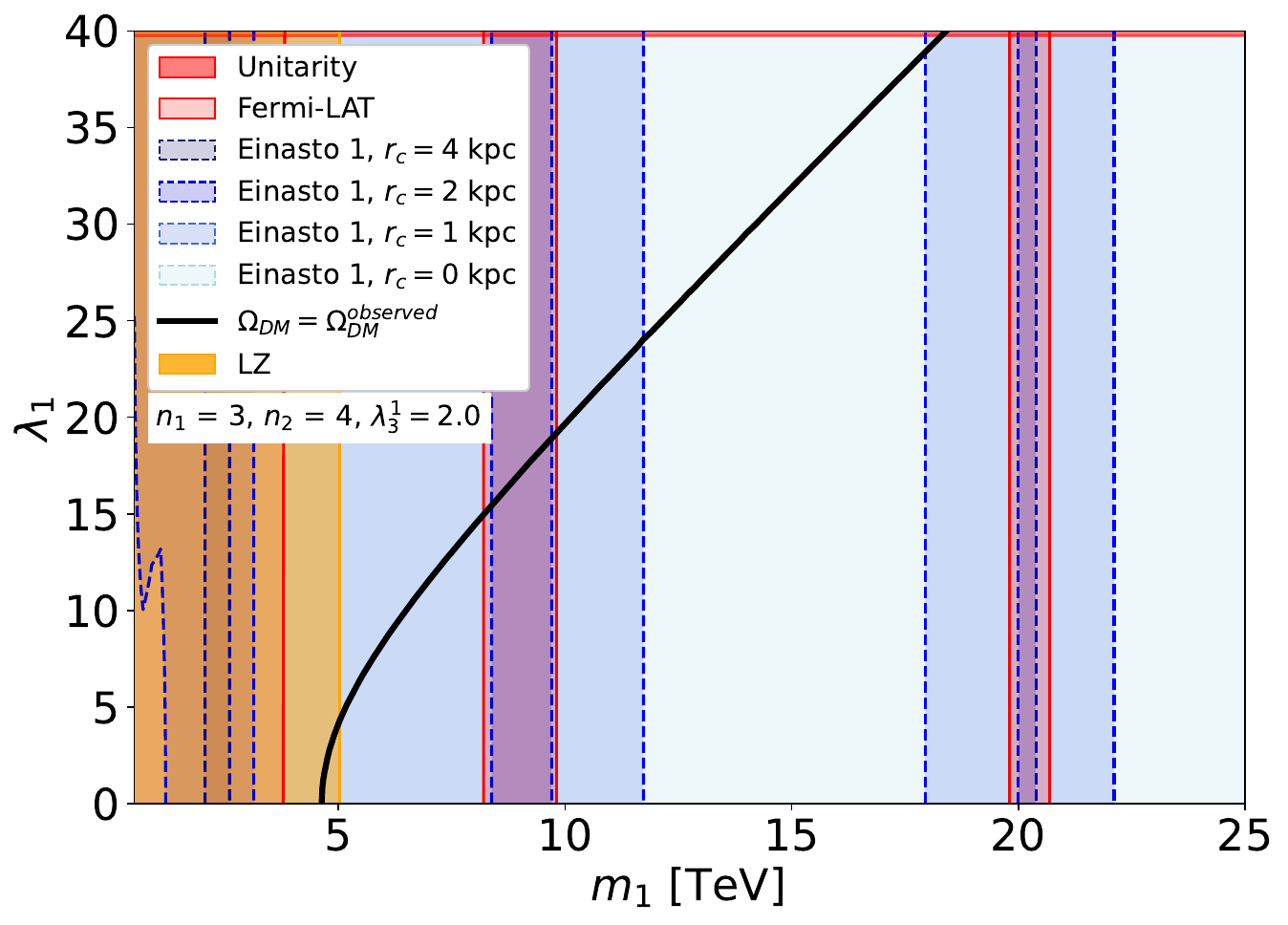}
        \label{fig:Scan3_34b}
    \end{subfigure}
    \begin{subfigure}{0.49\textwidth}
        \centering
        \caption{}
        \includegraphics[width=\textwidth]{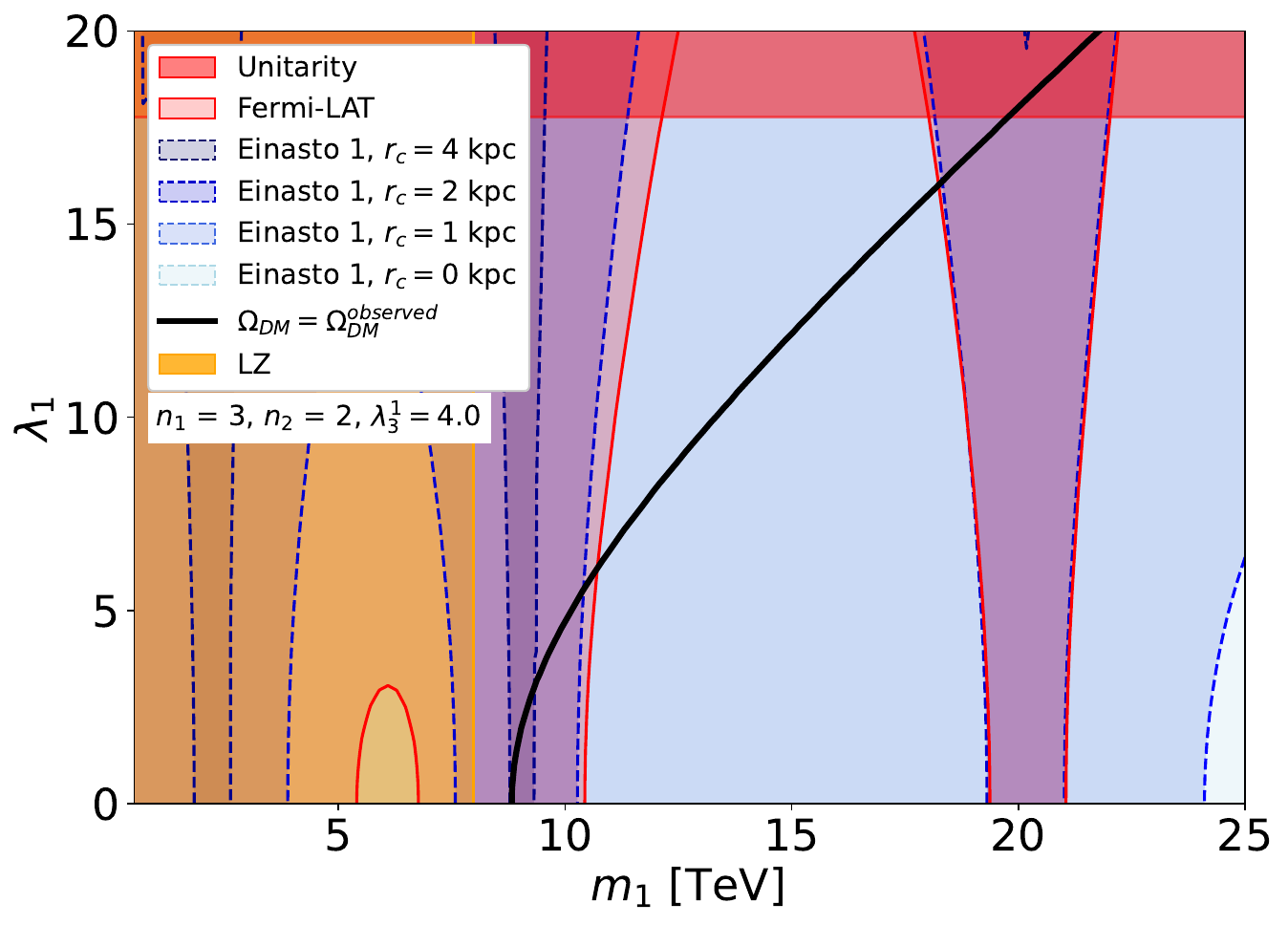}
        \label{fig:Scan3_32c}
    \end{subfigure}
    \begin{subfigure}{0.49\textwidth}
        \centering
        \caption{}
        \includegraphics[width=\textwidth]{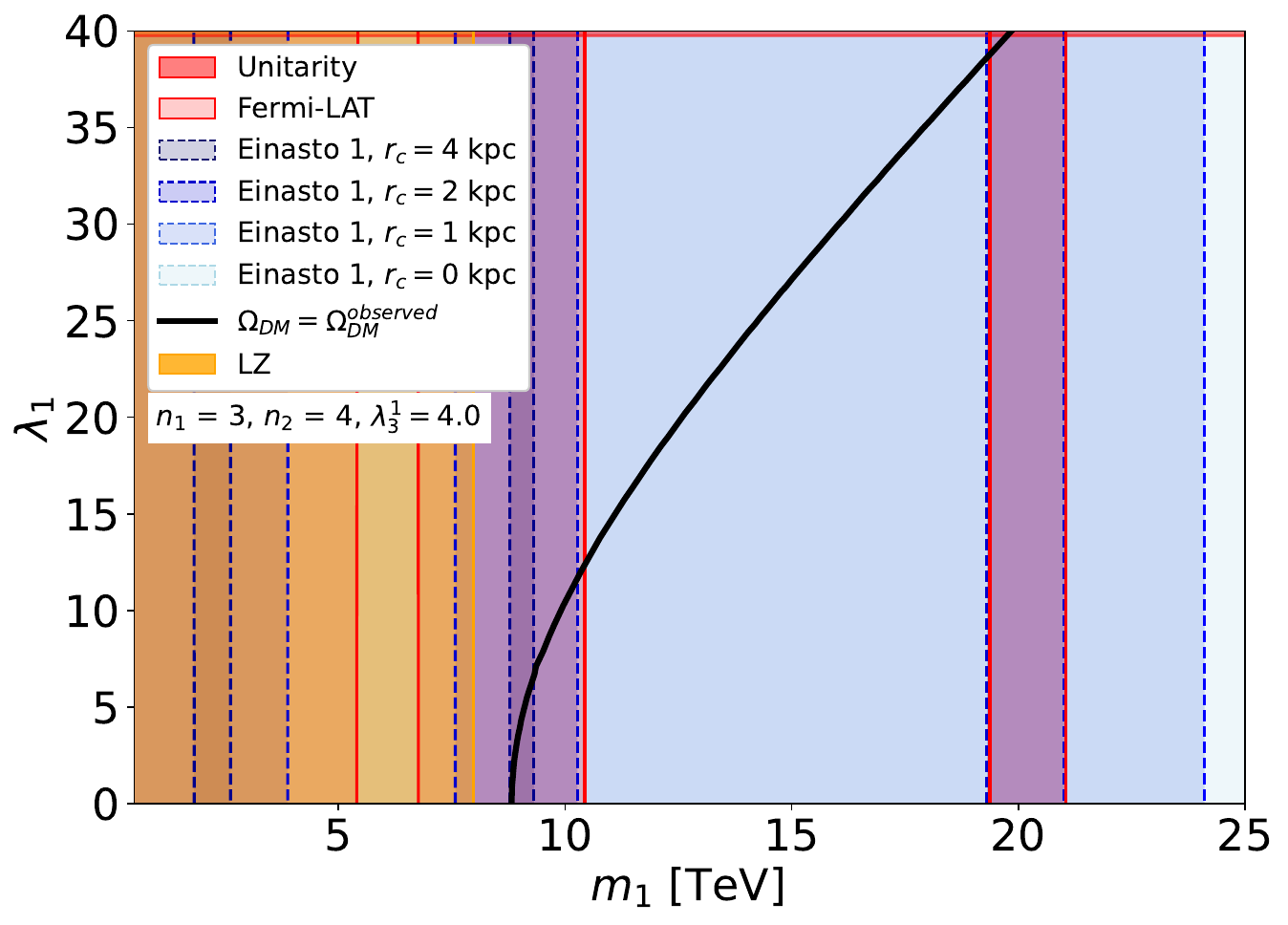}
        \label{fig:Scan3_34c}
    \end{subfigure}
    \caption{Similar to Fig.~\ref{fig:Scan1} but for different values of $\lambda_3^1$. The ratio $m_2/m_1 = 1.2$ and all other parameters are zero. Direct detection imposes limits in this case. The black line matches the observed dark matter abundance.}
    \label{fig:Scan3}
\end{figure}

Giving sizable values to the different coefficients of the $V_B$ part of the potential can modify the Sommerfeld enhancement and decrease the value of $\lambda_1$ necessary to reproduce the observed dark matter abundance. However, we find that an increase in the indirect detection signal generally overshadows any benefits. Direct detection can also become an important constraint in this case and limits are applied using the results of Appendix~\ref{Sec:DD}. The coupling of $V_B$ that can most significantly alter the dark matter abundance while avoiding constraints besides indirect detection is $\lambda_3^1$. Its impact is illustrated in Fig.~\ref{fig:Scan3}.

%%%%%%%%%%%%%%%%%%%%%%%%%%%%%%%%%%%%%%%%%%%%%%%%%%
\section{Conclusion}\label{Sec:Conclusion}
%%%%%%%%%%%%%%%%%%%%%%%%%%%%%%%%%%%%%%%%%%%%%%%%%%
The goal of this work is to study the constraints from indirect detection on inert multiplet semi-annihilation. For most multiplet sizes, the observed dark matter abundance can be obtained thermally, as long as the galactic dark matter density profile contains a sufficiently large core. For the case of a triplet and a quadruplet, semi-annihilation is Sommerfeld-suppressed and the correct abundance can be obtained even for a very cuspy dark matter density profile.

We end with a few concluding remarks. First, it has recently been claimed that the HESS experiment might have overestimated its effective area by a factor of $\sim 8$ \cite{Rodd:2024qsi}. If this indeed turns out to be true, our HESS limits would be considerably reduced, opening much of the parameter space. Since the claim of Ref.~\cite{Rodd:2024qsi} is rather extraordinary, we decided to take the conservative approach of assuming that the HESS results are correct and waiting for future validation of Ref.~\cite{Rodd:2024qsi}.

Second, there exist many subleading corrections that we did not include, as they were considered beyond the scope of a first study on indirect detection constraints on semi-annihilation of inert multiplets. This includes bound state formation, NLO corrections to the potential, NLO corrections to the cross sections and higher order corrections to the particle spectra (see e.g. Refs.~\cite{Mitridate:2017izz, Baumgart:2017nsr, Rinchiuso:2018ajn, Baumgart:2018yed, Beneke:2019qaa, Urban:2021cdu, Baumgart:2023pwn}). We leave such corrections for future work, but we do not expect any qualitative changes in our conclusions.

Third, the Cherenkov Telescope Array (CTA) is an array of upcoming imaging atmospheric Cherenkov telescopes similar to HESS, but with a larger effective area, a wider field of view and improved cosmic ray rejection~\cite{CTAConsortium:2017dvg}. As such, limits on the dark matter annihilation cross section in the galactic core are expected to improve in some cases by as much as an order of magnitude with respect to current HESS limits~\cite{CTA:2020qlo}. Such an improvement is likely to probe much of the remaining regions of parameter space where inert multiplet semi-annihilation can thermally explain the observed dark matter abundance.

\acknowledgments
This work was supported by the National Science and Technology Council under Grant No. NSTC-111-2112-M-002-018-MY3, the Ministry of Education (Higher Education Sprout Project NTU-113L104022-1), and the National Center for Theoretical Sciences of Taiwan.

\appendix

%%%%%%%%%%%%%%%%%%%%%%%%%%%%%%%%%%%%%%%%%%%%%%%%%%
\section{$SU(2)$ tensors}\label{Sec:SU2tensors}
%%%%%%%%%%%%%%%%%%%%%%%%%%%%%%%%%%%%%%%%%%%%%%%%%%
The $A_{abcd}$ tensor is
\begin{equation}\label{eq:Model2A}
  A_{abcd} = \sum_{M, M'=-J}^J C^{JM}_{j_1 m_1 j_2 m_2} C^{JM'}_{j_3 m_3 j_4 m_4} C^{00}_{J M J M'},
\end{equation}
where $C^{JM}_{j_1 m_1 j_2 m_2} = \langle j_1 j_2 m_1 m_2 |J M \rangle$ are the Clebsch-Gordan coefficients and
\begin{equation}\label{eq:Model2Ab}
  \begin{aligned}
    R   &= 4\left\lfloor\frac{n_2}{4}\right\rfloor  + 1, & J &= \frac{R - 1}{2},\\
    j_1 &= \frac{n_1 - 1}{2},      & j_2 &= \frac{n_1 - 1}{2},      & j_3 &= \frac{n_2 -1}{2},       & j_4 &= \frac{1}{2},\\ 
    m_1 &= \frac{n_1 + 1 - 2a}{2}, & m_2 &= \frac{n_1 + 1 - 2b}{2}, & m_3 &= \frac{n_2 + 1 - 2c}{2}, & m_4 &= \frac{3 - 2d}{2}.
  \end{aligned}
\end{equation}
The $B_{abc}$ tensor is
\begin{equation}\label{eq:Model2B}
  B_{abc} = \sum_{M=-J}^J C^{JM}_{j_1 m_1 j_2 m_2} C^{00}_{J M j_3 m_3},
\end{equation}
where
\begin{equation}\label{eq:Model2Bb}
  \begin{aligned}
    j_1 &= \frac{n_1 - 1}{2},      & j_2 &= \frac{n_1 - 1}{2},      & j_3 &= \frac{n_1 - 1}{2},      & J &= \frac{n_1 - 1}{2},\\ 
    m_1 &= \frac{n_1 + 1 - 2a}{2}, & m_2 &= \frac{n_1 + 1 - 2b}{2}, & m_3 &= \frac{n_1 + 1 - 2c}{2}.
  \end{aligned}
\end{equation}
The $C^r_{abcd}$ and $D^r_{abcd}$ tensors are written as
\begin{equation}\label{eq:Model2CD}
  \begin{aligned}
    C^1_{abcd} &= \delta_{ab} \delta_{cd}, \quad C^2_{abcd} = \tau^e_{ab} T^e_{cd},\\
    D^1_{abcd} &= \delta_{ab} \delta_{cd}, \quad D^2_{abcd} = \tau^e_{ab} \hat{T}^e_{cd},
  \end{aligned}
\end{equation}
where $\tau^e_{ab}$, $T^e_{cd}$ and $\hat{T}^e_{cd}$ are respectively the $SU(2)$ generators of dimension 2, $n_1$ and $n_2$. The $E_{abc}$ tensor is
\begin{equation}\label{eq:Model2E}
  E_{abc} = C^{JM}_{j_1 m_1 j_2 m_2},
\end{equation}
where
\begin{equation}\label{eq:Model2Eb}
  \begin{aligned}
    j_1 &= \frac{n_2 - 1}{2},      & j_2 &= \frac{1}{2},      & J &= \frac{n_1 - 1}{2},\\ 
    m_1 &= \frac{n_2 + 1 - 2b}{2}, & m_2 &= \frac{3 - 2c}{2}, & M &= \frac{n_1 + 1 - 2a}{2}.
  \end{aligned}
\end{equation}
The $F^r_{abcd}$ tensor is
\begin{equation}\label{eq:Model2F}
  F^r_{abcd} = \sum_{M=-J}^J C^{JM}_{j_1 m_1 j_2 m_2} C^{JM}_{j_3 m_3 j_4 m_4},
\end{equation}
where
\begin{equation}\label{eq:Model2Fb}
  \begin{aligned}
    J &= n_1 + 1 - 2r,\\
    j_1 &= \frac{n_1 - 1}{2},      & j_2 &= \frac{n_1 - 1}{2},      & j_3 &= \frac{n_1 -1}{2},       & j_4 &= \frac{n_1 -1}{2},\\ 
    m_1 &= \frac{n_1 + 1 - 2a}{2}, & m_2 &= \frac{n_1 + 1 - 2b}{2}, & m_3 &= \frac{n_1 + 1 - 2c}{2}, & m_4 &= \frac{n_1 + 1 - 2d}{2},
  \end{aligned}
\end{equation}
with $r \in \{1, 2, \ldots, \lfloor (n_1 + 1)/2 \rfloor\}$. The $G^r_{abcd}$ tensor is
\begin{equation}\label{eq:Model2G}
  G^r_{abcd} = \sum_{M=-J}^J C^{JM}_{j_1 m_1 j_2 m_2} C^{JM}_{j_3 m_3 j_4 m_4},
\end{equation}
where
\begin{equation}\label{eq:Model2Gb}
  \begin{aligned}
    J &= n_2 + 1 - 2r,\\
    j_1 &= \frac{n_2 - 1}{2},      & j_2 &= \frac{n_2 - 1}{2},      & j_3 &= \frac{n_2 -1}{2},       & j_4 &= \frac{n_2 -1}{2},\\ 
    m_1 &= \frac{n_2 + 1 - 2a}{2}, & m_2 &= \frac{n_2 + 1 - 2b}{2}, & m_3 &= \frac{n_2 + 1 - 2c}{2}, & m_4 &= \frac{n_2 + 1 - 2d}{2},
  \end{aligned}
\end{equation}
where $r \in \{1, 2, \ldots, \lfloor (n_2 + 1)/2 \rfloor\}$. Finally, the $H^r_{abcd}$ tensor is
\begin{equation}\label{eq:Model2H}
  H^r_{abcd} = \sum_{M=-J}^J C^{JM}_{j_1 m_1 j_2 m_2} C^{JM}_{j_3 m_3 j_4 m_4},
\end{equation}
where
\begin{equation}\label{eq:Model2Hb}
  \begin{aligned}
    J &= \frac{n_1 + n_2 - 2r}{2},\\
    j_1 &= \frac{n_1 - 1}{2},      & j_2 &= \frac{n_2 - 1}{2},      & j_3 &= \frac{n_1 -1}{2},       & j_4 &= \frac{n_2 -1}{2},\\ 
    m_1 &= \frac{n_1 + 1 - 2a}{2}, & m_2 &= \frac{n_2 + 1 - 2b}{2}, & m_3 &= \frac{n_1 + 1 - 2c}{2}, & m_4 &= \frac{n_2 + 1 - 2d}{2},
  \end{aligned}
\end{equation}
with $r \in \{1, 2, \ldots, \text{min}(n_1, n_2) \}$.

%%%%%%%%%%%%%%%%%%%%%%%%%%%%%%%%%%%%%%%%%%%%%%%%%%
\section{Direct detection}\label{Sec:DD}
%%%%%%%%%%%%%%%%%%%%%%%%%%%%%%%%%%%%%%%%%%%%%%%%%%
In the notation of Ref.~\cite{Hill:2014yxa}, the spin-independent per-nucleon cross section is
\begin{equation}\label{eq:DDCS}
  \sigma_{\text{SI}, N} = \frac{m_N^2 \hat{m}_0^2}{4\pi(m_N +  \hat{m}_0)^2}\left[\sum_q \frac{2 g g_V^q}{c_W m_Z^2} \hat{C}_{nn} F_1^{(N, q)}(0) - \frac{m_N}{m_h^2 \hat{m}_0}\frac{\hat{\Omega}_{nn}}{v_H} f^{(0)}_{q, N} \right]^2,
\end{equation}
where $n$ corresponds to the index of the lightest component, $N$ is either a proton or a neutron, $v_H \approx 246$~GeV is the Higgs vacuum expectation value and $g_V^q = T^3_q/2 - Q_q s_W^2$. For the $Z$ exchange, we use the values $F_1^{(p, u)}(0) = F_1^{(n, d)}(0) = 2$, $F_1^{(p, d)}(0) = F_1^{(n, u)}(0) = 1$ and set all other $F_1^{(N, q)}(0)$ to 0. For the Higgs exchange, we use $\sum_q f^{(0)}_{q, p} = 0.301$ and $\sum_q f^{(0)}_{q, n} = 0.307$~\cite{Ellis:2018dmb}. The values of $\sigma_{\text{SI}, N}$ are very close for the proton and neutron and their average is taken. The results of the LZ experiment from Ref.~\cite{LZ:2022lsv} are used to constrain $\sigma_{\text{SI}, N}$.

\bibliography{biblio}
\bibliographystyle{utphys}

\end{document}